\pdfoutput=1
\documentclass[acmsmall,screen,nonacm]{acmart}

\usepackage{xspace}
\usepackage[utf8]{inputenc}
\usepackage[T1]{fontenc}
\usepackage{soulutf8}
\usepackage[english]{babel}
\usepackage{csquotes}
\usepackage{xparse}
  \NewDocumentCommand{\li}{v}{\textbf{\smaller\texttt{#1}}}
\usepackage[frozencache,cachedir=minted-cache]{minted}
\usepackage{tikz}
  \usetikzlibrary{calc}
  \usetikzlibrary{shadows,graphs}
\usepackage{pgfplots}
\usepackage{mdframed}
\usepackage{amsmath}
\usepackage{mathpartir}
\usepackage[many]{tcolorbox}
\usepackage{newunicodechar}
\definecolor{OliveGreen}{HTML}{6AAFA0}

\DeclareUnicodeCharacter{2254}{:=}
\DeclareUnicodeCharacter{261B}{>}
\newunicodechar{÷}{$\div$}
\newunicodechar{×}{$\times$}
\newunicodechar{≤}{$\leqslant$}
\newunicodechar{≥}{$\geqslant$}
\newunicodechar{→}{$\rightarrow$}
\newunicodechar{≠}{$\neq$}

\newcommand\fstar{F$^\star$\xspace}
\newcommand\catala{\textbf{\textsf{Catala}}\xspace}

\newcommand{\sref}[1]{Section~\ref{sec:#1}}
\newcommand{\fref}[1]{Figure~\ref{fig:#1}}
\newcommand{\tref}[1]{Table~\ref{table:#1}}

\begin{document}

\title{Catala: A Programming Language for the Law}

\author{Denis Merigoux}
\email{denis.merigoux@inria.fr}
\orcid{0000-0003-2247-0938}
\affiliation{\institution{Inria}\country{France}}

\author{Nicolas Chataing}
\email{nicolas.chataing@ens.fr}
\affiliation{\institution{Inria, ENS Paris}\country{France}}

\author{Jonathan Protzenko}
\affiliation{\institution{Microsoft Research}\country{USA}}
\email{protz@microsoft.com}

\begin{abstract}
  Law at large underpins modern society, codifying and governing many aspects of
  citizens' daily lives. Oftentimes, law is subject to interpretation, debate
  and challenges throughout various courts and jurisdictions. But in some other
  areas, law leaves little room for interpretation, and essentially aims to
  rigorously describe a computation, a decision procedure or, simply said, an
  algorithm.

  Unfortunately, prose remains a woefully inadequate tool for the job. The lack
  of formalism leaves room for ambiguities; the structure of legal statutes,
  with many paragraphs and sub-sections spread across multiple pages, makes it
  hard to compute the intended outcome of the algorithm underlying a given text; and, as with any other
  piece of poorly-specified critical software, the use of informal, natural language
  leaves corner cases unaddressed.

  We introduce \catala, a new programming language that we specifically designed
  to allow a straightforward and systematic translation of statutory law into an
  executable implementation. Notably, \catala makes it natural and easy to express the
  general case / exceptions logic that permeates statutory law. \catala
  aims to bring together lawyers and programmers through a shared medium, which
  together they can understand, edit and evolve,
   bridging a gap that too often results in dramatically
  incorrect implementations of the law. We have implemented a compiler for
  \catala, and have proven the correctness of its core compilation steps using the
  \fstar proof assistant.

  We evaluate \catala on several legal texts that are algorithms in disguise,
  notably section 121 of the US federal income tax and the byzantine French
  family benefits; in doing so, we uncover a bug in the official implementation of the
  French benefits. We observe as a consequence of the formalization process
  that using \catala enables rich interactions between lawyers and
  programmers, leading to a greater understanding of the original legislative
  intent, while producing a correct-by-construction executable specification
  reusable by the greater software ecosystem. Doing so, \catala increases trust in legal
  institutions, and mitigates the risk of societal damage due to incorrect
  implementations of the law.

\end{abstract}



\maketitle

\section{Introduction}

We now know that since at least 2000 B.C.E. and the Code of
Ur-Nammu~\cite{urnammu}, various
societies have attempted to edict, codify and record their governing principles,
customs and rules in a set of legal texts -- the law. Nowadays, most aspects of
one's daily life are regulated by a set of laws or another, ranging from family
law, tax law, criminal law, to maritime laws, business laws or civil rights law.
No law is set in stone; laws are, over time, amended, revoked and modified by legislative bodies.
The resulting legal texts eventually reflect the complexity
of the process and embody the centuries of practice, debates, power struggles
and political compromises between various parties.

The practice of law thus oftentimes requires substantial human input. First, to
navigate the patchwork of exceptions, amendments, statutes and jurisprudence
relevant to a given case. Second, to fully
appreciate and identify the situation at play; understand whether one party
falls in a given category or another; and generally classify and categorize, in
order to interpret a real-world situation into something the law can talk about.

This latter aspect is perhaps the greatest challenge for a computer
scientist: a general classification system remains an elusive prospect when
so much human judgement and appreciation is involved. Fortunately, a subset of
the law, called \emph{computational law} or sometimes \emph{rules as code}, concerns itself with situations
where entities are well-defined, and where human appreciation, judgement or
interpretation are not generally expected. Examples of computational law include,
but are not limited to: tax law, family benefits, pension computations, monetary
penalties and private billing contracts. All of these are algorithms in
disguise: the law (roughly) defines a function that produces outcomes based on a
set of inputs.

As such, one might think computational law would be easily translatable into a computer program.
Unfortunately, formidable challenges remain.
First, as mentioned above, the law is the result of a centuries-long process:
its convoluted structure demands tremendous expertise and training to
successfully navigate and understand, something that a computer programmer may
not have. Second, the language in which legal statutes are drafted
is so different from existing
programming languages that a tremendous gap remains between the legal text and
its implementation, leaving the door open for discrepancies, divergence and
eventual bugs, all with dramatic societal consequences.

Examples abound. In France, the military's payroll computation involves 174
different bonuses and supplemental compensations. Three successive attempts were
made to rewrite and modernize the military paycheck infrastructure; but with a
complete disconnect between the military statutes and the implementation teams
that were contracted, the system had to be scrapped~\cite{louvois}. Software
engineering failures are perhaps a fact of life in the IT world; but in this
particular case, actual humans bore the consequences of the
failure, with military spouses
receiving a 3-cent paycheck, or learning years later that
they owe astronomical amounts to the French state.
Perhaps more relevant to the current news, the US government
issued a decree intended to provide financial relief to US taxpayers whose
personal economic situation had been affected by the Covid-19 pandemic. Owing to
an incorrect implementation by the Internal Revenue Service (IRS), nearly one
million Americans received an incorrect Economic Impact Payment (EIP), or none
at all~\cite{gao-eip}.

Both examples are similar, in that a seemingly pure engineering failure turns
out to have dramatic consequences in terms of human livelihoods. When a family
is at the mercy of the next paycheck or EIP, a bug in those systems could mean
bankruptcy. In our view, these is no doubt that these systems are yet another
flavor of critical software.

A natural thought is perhaps to try to simplify the law itself.
Unfortunately, this recurrent theme of the political discourse
often conflicts with the political reality that requires compromise and
fined-grained distinctions. Hence, the authors do not anticipate a drastic improvement around
the world concerning legal complexity.
Therefore, our only hope for improvement lies on the side of programming
languages and tools.

Tax law provides a quintessential example. While many of the implementations
around the world are shrouded in secrecy, the public occasionally gets a glimpse
of the underlying infrastructure. Recently, Merigoux \emph{et al.}~\cite{mlang}
reverse-engineered the computation of the French income tax, only to discover
that the tax returns of an entire nation were processed using an antiquated system
designed in 1990, relying on 80,000 lines of code written in a custom, in-house
language, along with 6,000 lines of hand-written C directly manipulating tens of
thousands of global variables. This particular situation highlights the perils
of using the wrong tool for the job: inability to evolve, resulting in
hand-written C patch-ups; exotic semantics which make reproducing the
computation extremely challenging; and a lack of accountability, as the cryptic
in-house language cannot be audited by anyone, except by the handful of experts
who maintain it. This is by no means a “French exception”: owing to an infrastructure
designed while Kennedy was still president, the IRS recently handed over
\$300,000,000's worth of fraudulent refunds to taxpayers~\cite{irs-60}. The
rewrite, decades in planning, keeps being pushed further away in the
future~\cite{irs-60-2}.

In this work, we propose a new language, \catala, tailored specifically for the
purpose of faithfully, crisply translating computational law into executable
specifications. \catala is designed to follow the existing structure of legal
statutes, enabling a one-to-one correspondence between a legal paragraph and
its corresponding translation in \catala. Under the hood, \catala uses
prioritized default logic~\cite{brewka2000}; to the best of our knowledge,
\catala is the first instance of a programming language designed with this logic
as its core system.
\catala has clear semantics, and
compiles to a generic lambda-calculus that can then be translated to any
existing language. We formalize the compilation scheme of \catala with \fstar
and show that it is correct. In doing so, we bridge the gap between legal
statutes and their implementation; we avoid the in-house language trap; and we
provide a solid theoretical foundation to audit, reproduce, evaluate and
reuse computational parts of the law.
Our evaluation, which includes user studies, shows that \catala can successfully
express complex sections of the US Internal Revenue Code and the French family
benefits computation.

The benefits of using \catala are many: lawmakers and lawyers are given a formal language that
accurately captures their intent and faithfully mirrors the prose; programmers
can derive and distribute a canonical implementation, compiled to the programming
language of their choice; citizens can audit, understand and evaluate
computational parts of the law; and advocacy groups can shed more light on
oftentimes obscure, yet essential, parts of civil society.

\section{Background on Legal Texts \& their Formalization}
\label{sec:background}

Legal statutes are written in a style that can be confounding for a computer
scientist. While a program's control-flow (as a first approximation) makes
forward progress, statutes frequently back-patch previous definitions and
re-interpret earlier paragraphs within different contexts. The result more
closely resembles assembly with arbitrary jumps and code rewriting, rather than
a structured language.

To illustrate how the law works, we focus on Section 121 of the US Internal Revenue
Code~\cite{section-121}, our running example throughout this paper. Section 121
is written in English, making it accessible to an international audience without
awkward translations; it features every single difficulty we wish to tackle with
\catala; and it is a well-studied and well-understood part of the tax law. We go
through the first few paragraphs of the section; for each of them, we informally
describe the intended meaning, then highlight the surprising semantics of the
excerpt. These paragraphs are contiguous in the law;
we intersperse our own commentary in-between the quoted blocks.

\subsection{Overview of Section 121}
\label{sec:121a}

Section 121 is concerned with the “Exclusion of gain from sale of principal
residence”. In common parlance, if the taxpayer sells %
their residence, they are not required to report the profits as income, hence
making such profits non-taxable. Paragraph (a) defines the exclusion itself.

\mdfsetup{
    backgroundcolor=gray!10,
    linewidth=0pt,
    skipabove=.5em,
    skipbelow=.5em,
    leftmargin=2em,
    rightmargin=2em,
    usetwoside=false,
}
\newcommand\myquote[1]{
  \begin{mdframed}
#1
  \end{mdframed}
}

\myquote{
\textbf{(a) Exclusion} \\
  Gross income shall not include gain from the sale or exchange of property if,
  during the 5-year period ending on the date of the sale or exchange, such
  property has been owned and used by the taxpayer as the taxpayer’s principal
  residence for periods aggregating 2 years or more.
}

\noindent
The part of the sentence that follows the “if” enumerates conditions under which
this tax exclusion can be applied. This whole paragraph is valid \emph{unless
specified otherwise}, as we shall see shortly.

\subsection{Out-of-order Definitions}
\label{sec:121b}

Paragraph (b) immediately proceeds to list \emph{limitations}, that is,
situations in which (a) does not apply, or needs to be refined. Section 121 thus
consists of a general case, (a), followed by a long enumeration of limitations
ranging from (b) to (g). We focus only on (b). The first limitation (b)(1) sets a
maximum for the exclusion, “generally” \$250,000. Left implicit is the fact that
any proceeds of the sale beyond that are taxed as regular income.

\myquote{
\textbf{(b) Limitations} \\
\textbf{(1) In general} \\
The amount of gain excluded from gross income under subsection (a) with respect
to any sale or exchange shall not exceed \$250,000.
}

\noindent
We remark that even though (b)(1) is a key piece of information for the
application of Section 121, the reader will find it only if they keep going
after (a).
This is a general feature of legal texts: relevant
information is scattered throughout, and (a) alone is nowhere near enough
information to make a determination of whether the exclusion applies to a
taxpayer.

\subsection{Backpatching; Exceptions}
\label{sec:121b1A}

Entering (b)(2), paragraph (A) \emph{modifies} (b)(1) \emph{in place}, stating
for “joint returns” (i.e. married couples), the maximum exclusion can be \$500,000.

\myquote{
\textbf{(A) \$500,000 Limitation for certain joint returns} \\
  Paragraph (1) shall be applied by substituting “\$500,000” for “\$250,000” if— \\
  (i) either spouse meets the ownership requirements of subsection (a) with
  respect to such property; \\
  (ii) both spouses meet the use requirements of subsection (a) with respect to
  such property; and \\
  (iii) neither spouse is ineligible for the benefits of subsection (a) with
  respect to such property by reason of paragraph (3).
  }

\noindent
Several key aspects of paragraph (A) are worth mentioning. First,
(A) \emph{backpatches} paragraph (b)(1); the law essentially encodes a
search-and-replace in its semantics.

Second, (A) \emph{overrides} a previous
general case under specific conditions.
In a functional programming language, a
pattern-match first lists the most specific matching cases, and catches all
remaining cases with a final wildcard. A text of law proceeds in the exact
opposite way: the general case in (a) above is listed first, then followed by
limitations that modify the general case under certain conditions. This is by
design: legal statutes routinely follow a “general case first, special cases later”
approach which mirrors the legislator's intentions.

Third, conditions (i) through (iii) are a conjunction, as indicated by
the \enquote{and} at the end of (ii). We note that (iii)
contains a forward-reference to (3) which we have not seen yet.
(Through our work, we fortunately have never encountered a circular reference.)

\subsection{Re-interpreting}

If limitation (A) doesn't apply, we move on to (B), which essentially stipulates
that the exclusion in (b)(1) should be re-interpreted for each spouse separately
\emph{as if they were not married}; the final exclusion is then the sum of the
two sub-computations.

\myquote{
\textbf{(B) Other joint returns} \\
If such spouses do not meet the requirements of subparagraph (A), the limitation
under paragraph (1) shall be the sum of the limitations under paragraph (1) to
which each spouse would be entitled if such spouses had not been married. For
purposes of the preceding sentence, each spouse shall be treated as owning the
property during the period that either spouse owned the property.
}

\noindent
We thus observe that the law is re-entrant and can call itself recursively under different
conditions. This is indicated here by the use of the conditional tense, i.e.
“would”.

\subsection{Out-of-order Backpatching}

In another striking example, (3) cancels the whole exclusion (a) under
certain conditions.

\myquote{
\textbf{(3) Application to only 1 sale or exchange every 2 years}\\
Subsection (a) shall not apply to any sale or exchange by the taxpayer if,
during the 2-year period ending on the date of such sale or exchange, there was
any other sale or exchange by the taxpayer to which subsection (a) applied.
}

\noindent
Paragraph (3) comes a little further down; a key takeaway is that, for a
piece of law, one must process the \emph{entire} document; barring that,
the reader might be missing a crucial limitation that only surfaces much later
in the text.

\subsection{A Final Example}

Paragraph (4) concerns the specific case of a surviving spouse that sells
the residence within two years of the death of their spouse, knowing that the
conditions from (A) applied (i.e. “returned true”) \emph{right before the date
of the death}.

\myquote{
\textbf{(4) Special rule for certain sales by surviving spouses}\\
In the case of a sale or exchange of property by an unmarried individual whose
spouse is deceased on the date of such sale, paragraph (1) shall be applied by
substituting “\$500,000” for “\$250,000” if such sale occurs not later than 2
years after the date of death of such spouse and the requirements of paragraph
(2)(A) were met immediately before such date of death.
}

\noindent
Paragraph (4) combines several of the complexities we saw above. It not only
back-patches (1), but also recursively calls (2)(A) under a different
context, namely, executing (2)(A) at a \emph{previous date} in
which the situation was different.
In functional programming lingo, one might say that there is a hidden lambda in
(2)(A), that binds the date of the sale.

\subsection{Formal Insights on Legal Logic}
\label{sec:insights}

We have now seen how legal statutes are written, the thought
process they exhibit, and how one is generally supposed to interpret them. We wish
to emphasize that the concepts described are by no means specific to tax law or
the US legal system: we found the exact same patterns in other parts of US law
and non-US legal systems. Section 121 contains many more paragraphs; however,
the first few we saw above are sufficient to illustrate the challenges in formally
describing the law.

The main issue in modeling legal texts therefore lies in their underlying logic,
which relies heavily on the pattern of having a default case followed by
exceptions.  This nonmonotonic logic is known as \emph{default
logic}~\cite{REITER198081}.  Several refinements of default logic have been
proposed over time; the one closest to the purposes of the law is known as
prioritized default logic~\cite{brewka2000}, wherein default values are guarded
by justifications, and defaults can be ordered according to their relative
precedence. Lawsky~\cite{lawsky2018} argues that this flavor of default logic
is the best suited to expressing the law. We concur, and adopt prioritized
default logic as a foundation for \catala.

In default logic, formulas include defaults, of the form $a:\vec{b_i}/c$, wherein: if
formula $a$ holds; if the $\vec{b_i}$ are consistent with the set of known facts; then
$c$ holds. One can think of $a$ as the precondition for $c$, and the $\vec{b_i}$
as a set of exceptions that will prevent the default fact $c$ from being
applicable.
Prioritized logic adds a strict partial order over defaults, to
resolve conflicts when multiple defaults may be admissible at the same time.

The main design goal of \catala is to provide the design and implementation of a
language tailored for the law, using default logic as its core building block,
both in its syntax and semantics. \catala thus allows lawyers to express the
general case / exceptions pattern naturally. We now informally present \catala.

\section{A \catala Tutorial}
\label{sec:tutorial}

Our introduction to legal texts in \sref{background} mixes informal,
high-level overviews of what each paragraph intends to express, along with excerpts
from the law itself. Our English prose is too informal to express anything
precisely; but “legalese” requires a high degree of familiarity with the law to
successfully grasp all of the limitations and compute what may or may not be
applicable to a given taxpayer's situation.

We now introduce \catala by example, and show how the subtleties of each
paragraph can be handled unambiguously and clearly by \catala. Our guiding
principle is twofold: we want to formally express the intended meaning without
being obscured by the verbosity of legal prose; yet, we wish to remain close to
the legal text, so that the formal specification remains in close correspondence
with the reference legal text, and can be understood by lawyers. \catala
achieves this with a dedicated surface language that allows legal experts to
follow their usual thinking.

\subsection{Metadata: Turning Implicits Into Explicits}
\label{sec:metadata}

Legal prose is very dense, and uses a number of concepts without explicitly defining them
in the text. For instance, in Section 121, the notion of time period is
implicit, and so are the various types of tax returns one might file (individual
or joint). Furthermore, entities such as the taxpayers (whom we will call
“Person 1” and “Person 2”) need to be materialized. Finally, for each one of
those entities, there are a number of inputs that are implicitly referred to
throughout the legal statute, such as: time periods in which each Person was
occupying the residence as their primary home; whether there was already a sale
in the past two years; and many more, as evidenced by the myriad of variables
involved in (i)-(iii).

Our first task when transcribing legal prose into a formal \catala description
is thus to enumerate all structures, entities and variables relevant to the
problem at stake.
We provide the definitions and relationships between variables later on.
This is a conscious design choice of \catala: before even talking about
\emph{how} things are computed, we state \emph{what} we are talking about. In
doing so, we mimic the behavior of lawyers, who are able to infer what
information is relevant based on the legal text. We call this description of
entities the \emph{metadata}.

\newcommand*\FancyVerbStartString{```catala}
\newcommand*\FancyVerbStopString{```}

\newcommand\lref[1]{\ref{line:#1}}

\begin{minted}[xleftmargin=2em,linenos,escapeinside=!!,firstnumber=1,fontsize=\small]{catala_en}
```catala
declaration structure Period:
  data start content date
  data end content date

declaration structure PersonalData:
  data property_ownership content collection Period
  data property_usage_as_principal_residence content collection Period

declaration scope Section121SinglePerson:!\label{line:single}!
  context gain_from_sale_or_exchange_of_property content money!\label{line:single:gain}!
  context personal content PersonalData
  context requirements_ownership_met condition!\label{line:single:ownership}!
  context requirements_usage_met condition!\label{line:single:usage}!
  context requirements_met condition!\label{line:single:met}!
  context amount_excluded_from_gross_income_uncapped content money!\label{line:single:uncapped}!
  context amount_excluded_from_gross_income content money!\label{line:single:gross}!

  context aggregate_periods_from_last_five_years content duration!\label{line:helper}!
    depends on collection Period
```
\end{minted}

\noindent
\catala
features a number of built-in types. \li+date+s are triples of a year,
month and a day. \catala provide syntax for US-centric and non-US-centric input
formats. Distinct from \li+date+ is \li+duration+, the type of a time interval,
always expressed as a number of days. If the law introduces durations such as
"two years", it is up to the user to specify how "two years" should be
interpreted. Booleans have type \li+condition+. Monetary amounts have type
\li+money+. The higher-kinded type \li+collection+ is also built-in.

The snippet above shows an excerpt from Section 121's metadata. The first two
\li+declaration+s declare product types via the \li+structure+ keyword.  The
type \li+Period+ contains two fields, \li+start+ and \li+end+.

A word about aesthetics: while programmers generally prize compactness of
notation, advocating e.g. point-free-styles or custom operators, non-experts
are for the most part puzzled by compact notations. Our surface syntax
was designed in collaboration with lawyers, who confirmed that the generous
keywords improve readability, thus making \catala easier to understand by legal
minds.

Line~\lref{single} declares \li+Section121SinglePerson+, a \li+scope+.  A
key technical device and contribution of \catala, scopes allow the programmer to
follow the law's structure,
revealing the implicit modularity in legal texts.
Scopes are declared in the metadata section: the \li+context+ keyword indicates
that the value of the field might be determined later, \emph{depending on the
context}.
Anticipating on \sref{formalizing}, the intuition is that \li+scope+s are
functions and \li+context+s are their parameters and local variables.

Context variables are declarative; once a \catala program is compiled to a
suitable target language, it is up to the programmer to invoke a given scope
with suitable initial values for those context variables that are known to be
inputs of the program; after the \catala program has executed, the programmer
can read from the context variables that they know are outputs of the program.
From the point of view of \catala, there is no difference between input and
output variables; but we are planning a minor syntactic improvement to allow
programmers to annotate these variables for readability and simpler
interoperation with hand-written code. If at run-time the program reads the
value of a context variable that was left unspecified by the programmer,
execution aborts.

The main purpose of Section 121 is to talk about the gain that a person derived from
the sale of their residence (line \lref{single:gain}), of type \li+money+. Paragraph (a)
implicitly assumes the existence of time periods of ownership and usage of the
residence; we materialize these via the \li+personal+ field which holds two
\li+collection Period+s.
These in turn allow us to define whether the ownership
and usage requirements are met (of type \li+condition+, lines
\lref{single:ownership}-\lref{single:usage}). A further
condition captures whether \emph{all} requirements are met (line
\lref{single:met}).
The outcome of the law is the amount that can be excluded from the gross income,
of type \li+money+ (line \lref{single:gross}). (The need for an intermediary
variable at line \lref{single:uncapped}
becomes apparent in \sref{tuto-limitations}.)
A local helper computes the aggregate number of days in a set
of time periods; the helper takes a single argument of type
\li+collection Period+ (line \lref{helper}) and, being a local closure,
can capture other \li+context+ variables.

\subsection{Scopes and Contexts: Declarative Rules and Definitions}
\label{sec:rulesdefinitions}
We now continue with our formalization of (a) and define the
context-dependent variables, as well as the relationships between them. \catala
is declarative: the user relates \li+context+ variables together, via the
\li+definition+ keyword, or the \li+rule+ keyword for
\li+condition+s. We offer separate syntax for two reasons. First, for legal
minds, conditions and data are different objects and reflecting that in the
surface syntax helps with readability. Second, there is a core semantic difference:
booleans (conditions) are false by default in the law; however, other types of
data have no default value. Internally, \catala desugars \li+rule+s to
\li+definition+s equipped with a default value of \li+false+ (\S\ref{sec:desugaring}).

\begin{minted}[xleftmargin=2em,escapeinside=!!,linenos,firstnumber=1,fontsize=\small]{catala_en}
```catala
scope Section121SinglePerson:
  rule requirements_ownership_met under condition
    aggregate_periods_from_last_five_years of personal.property_ownership >=^
      730 day
  consequence fulfilled
  rule requirements_usage_met under condition
    aggregate_periods_from_last_five_years of
      personal.property_usage_as_principal_residence >=^ 730 day
  consequence fulfilled
  rule requirements_met under condition
    requirements_ownership_met and requirements_usage_met
  consequence fulfilled
  definition amount_excluded_from_gross_income_uncapped equals
    if requirements_met then gain_from_sale_or_exchange_of_property else $0
```
\end{minted}

\noindent
Lines 2-4 capture the ownership requirement, by calling the helper
\li+aggregate_periods_...+ with argument
\li+property_ownership+, a previously-defined context variable. (The full
definition of the helper, which involves another
context variable for the date of sale, is available in
the artifact \cite{merigoux_denis_2021_4775161}.) Paragraph (a) states “for periods aggregating 2 years or more”:
for the purposes of Section 121, and as defined in Regulation 1.121-1(c)(1), a
year is always 365 days. \catala resolves the ambiguity by simply not offering
any built-in notion of yearly duration, and thus makes the law clearer. The
\li+^+ suffix of the comparison operator \li+>=^+ means that we are comparing
durations.

The ownership requirement is “fulfilled” (i.e. defined to \li+true+) under
a certain condition. This is our first taste of prioritized default logic
expressed through the syntax of \catala: the built-in default, set to
\li+false+, is overridden with a rule that has higher priority. This is a simple
case and more complex priorities appear in later sections. However, this
example points to a key insight of \catala: rather than having an arbitrary
priority order resolved at run-time between various \li+rule+s,
we encode priorities statically in the surface syntax
of the language, and the pre-order is derived directly from the syntax tree of
rules and definitions. We explain this in depth later on (\sref{desugaring}).

Similarly, lines 5-7 define the usage requirement using the \li+rule+ keyword to trigger a
\li+condition+: the value of
\li+requirements_usage_met+ is \li+false+ unless the boolean
expression at lines 6-7 is \li+true+. One legal subtlety, implicit in
(a), is that ownership and usage periods do not have to overlap. The \catala program makes
this explicit by having two collections of time periods.

The requirements are met if both ownership and usage requirements are met (lines
9-11). In that case, the income gain can be excluded from the income tax (lines
12-13). The latter is defined via the \li+definition+ keyword, as \li+rule+ is
reserved for booleans.

We have now formalized Paragraph (a) in \catala. At this stage, if the user
fills out the remaining inputs, such as the gain obtained from the sale of the
property, and the various time periods, the interpreter automatically computes the
resulting value for the amount to be excluded from the gross income.
The interpreter does so by performing a control-flow analysis
and computing a topological sort of assignments. Cycles are
rejected, since the law is not supposed to have dependency cycles.
(\sref{formalizing} describes the full semantics of the language.)

We note that a single sentence
required us to explicitly declare otherwise implicit concepts, such as the
definition of a year; and to clarify ambiguities, such as whether time periods may
overlap. With this concise example, we observe that the benefits of formalizing
a piece of law are the same as formalizing any piece of critical software:
numerous subtleties are resolved, and non-experts are provided with an explicit,
transparent executable specification that obviates the need for an expert legal
interpretation of implicit semantics.

\subsection{Split Scopes: Embracing the Structure of the Law}
\label{sec:tuto-limitations}

We now move on to limitations (\S\ref{sec:121b}).
A key feature of \catala is that it enables a
literate programming style~\cite{knuthliterateprogramming}, where each paragraph of law is immediately followed
by its \catala transcription.
Now that we're done with (a), we insert a textual
copy of the legal prose for (b), then proceed to transcribe it in \catala.

\begin{minted}[xleftmargin=2em,linenos,firstnumber=1,fontsize=\small]{catala_en}
```catala
scope Section121SinglePerson:
  definition gain_cap equals $250,000
  definition amount_excluded_from_gross_income equals
    if amount_excluded_from_gross_income_uncapped >=$ gain_cap then gain_cap
    else amount_excluded_from_gross_income_uncapped
```
\end{minted}

\noindent
In Paragraph (b)(1), the law overwrites the earlier definition from (a) and
re-defines it to be capped by \$250,000. In line with our earlier
design choices, we rule out confusion and rely on the auxiliary variable (the
“uncapped” variant), to then compute the final amount excluded from the gross
income (lines 4-8). Out-of-order definitions that are provided at a later phase
in the source are an idiomatic pattern in \catala.

\subsection{Complex Usage of the Default Calculus; Exceptions}
\label{sec:exceptions}

Before making any further progress, we need to introduce new entities to take
into account the fact that we may now possibly be dealing with a joint return.
We introduce a new abstraction or, in \catala lingo, scope:
\li+Section121Return+.

\begin{minted}[xleftmargin=2em,linenos,firstnumber=1,fontsize=\small]{catala_en}
```catala
declaration structure CoupleData:
  data personal1 content PersonalData
  data personal2 content PersonalData

declaration enumeration ReturnType:
  -- SingleReturn content PersonalData
  -- JointReturn content CoupleData

declaration scope Section121Return:
  context return_data content ReturnType
  context person1 scope Section121SinglePerson
  context person2 scope Section121SinglePerson
  context gain_cap content money
```
\end{minted}

\noindent
We follow the out-of-order structure of the law; only from here on do we consider
the possibility of a joint return. Having introduced a new level of abstraction,
we need to relate the \li+ReturnType+ to the persons involved. We do so by
introducing new equalities, of which we show the first one.

\begin{minted}[xleftmargin=2em,linenos,firstnumber=1,fontsize=\small]{catala_en}
```catala
scope Section121Return:
  definition person1.personal equals match return_data with
  -- SingleReturn of personal1 : personal1
  -- JointReturn of couple : couple.personal1
```
\end{minted}

\noindent
Having set up a proper notion of joint return, we now turn our attention to
(b)(2)(A) (\S\ref{sec:121b1A}).

\begin{minted}[xleftmargin=2em,linenos,firstnumber=1,fontsize=\small,escapeinside=!!]{catala_en}
```catala
scope Section121Return:
  definition gain_cap equals person1.gain_cap!\label{line:joint:gain}!
  rule paragraph_A_applies under condition!\label{line:joint:applies}!
    (return_data is JointReturn) and
    (person1.requirements_ownership_met or person2.requirements_ownership_met) and
    (person1.requirements_usage_met and person2.requirements_usage_met) and
    (not (paragraph_3_applies of person1.other_section_121a_sale)) and!\label{line:joint:helper}!
    (not (paragraph_3_applies of person2.other_section_121a_sale))
  consequence fulfilled
  exception!\label{line:joint:exception}! definition gain_cap under condition paragraph_A_applies
    consequence equals $500,000
```
\end{minted}

\noindent
Until now, the gain cap was defined to be that of the taxpayer, that is, Person
1 (line~\lref{joint:gain}). We now need to determine whether the conditions from
Paragraph (A) apply (line~\lref{joint:applies}). To that end, we introduce an
intermediary variable, \li+paragraph_A_applies+. This variable will be used
later on for (B), whose opening sentence is “if such spouses do not meet the
requirements of (A)”.

We now introduce the notion of \li+exception+ (line~\lref{joint:exception}). In
\catala, if, at run-time, more than a single applicable definition for any
\li+context+ variable applies, program execution aborts with a fatal error.
In the absence of the \li+exception+ keyword, and in the presence of a joint
return that satisfies paragraph (A), the program would be statically
accepted by \catala, but would be rejected at run-time: there are two
definitions for \li+gain_cap+, both their conditions hold (\li+true+ and
\li+paragraph_A_applies+), and there is no
priority ordering indicating how to resolve the conflict.
The \li+exception+ keyword allows solving this very issue. The keyword indicates
that, in the pre-order of definitions, the definition at
line~\lref{joint:exception} has a higher priority than the one at
\lref{joint:gain}.

Generally, \catala allows an arbitrary tree of definitions each refined
by exceptions, including exceptions to exceptions (which we have encountered in
the law); the rule of thumb remains: only one single definition should be
applicable at a time, and any violation of that rule indicates either programmer
error, or a fatal flaw in the law.

\subsection{Recapping}

Modeling the law is labor-intensive, owing to all of the implicit assumptions
present in what is seemingly “just legalese”. In our experience, this process is
best achieved through pair-programming, in which a \catala expert transcribes a
statute with the help of a legal expert.
We thus stop here our \catala tutorial and defer the full modelization of
\sref{background} to the artifact \cite{merigoux_denis_2021_4775161}.
Briefly, modeling (B) requires introducing a new scope
for a two-pass processing that models the re-entrancy (“if such spouses
had not been married”). Modeling the forward-reference to (3) requires
introducing a helper \li+paragraph_3_applies+ whose \li+definition+ is provided
later on, after Paragraph (3) has been suitably declared
(line~\lref{joint:helper}, above).

As this tutorial wraps up, we look back on all of the language features we
presented.
While \catala at first glance resembles a functional language with heavy syntactic
sugar, diving into the subtleties of the law exhibits the need for two features
that are not generally found in lambda-calculi. First, we allow the user to
define context variables through a combination of an (optional) default case,
along with an arbitrary number of special cases, either prioritized or
non-overlapping. The theoretical underpinning of this feature is the
\emph{prioritized default calculus}. Second, the
out-of-order nature of definitions means that \catala is entirely declarative,
and it is up to the \catala compiler to compute a suitable dependency order for
all the definitions in a given program.
Fortunately, the law does not have general recursion, meaning that we do not
need to compute fixed points, and do not risk running into circular definitions.
Hence, our language is not Turing-complete, purposefully.

We mentioned earlier (\sref{insights}) that we have found both US and French
legal systems to exhibit the same patterns in the way their statutes are
drafted. Namely, the general case / exceptions control-flow; out-of-order
declarations; and overrides of one scope into another seem to be universal
features found in all statutes, regardless of the country or even the language
they are drafted in. Based on conversations with S. Lawsky, and a broader
assessment of the legal landscape, we thus posit that \catala captures the
fundamentals of legal reasoning.

\section{Formalizing \catala}
\label{sec:formalizing}

We now formally introduce the semantics and compilation of \catala.
Notably, we focus on what makes \catala special: its default calculus.
To that end,
we describe a series of compilation steps: we desugar the concrete syntax to a
\emph{scope language}; we define the semantics of scopes via a translation to a
\emph{default calculus}; we then finally compile the \emph{default calculus} to
a language equipped with exceptions, such as OCaml. This last part is where the
most crucial compilation steps occur: we prove its soundness via a mechanization
in the \fstar proof assistant.

\subsection{From \catala to the Scope Language}
\label{sec:desugaring}

The scope language resembles \catala's user-facing language: the notion of scope
is still present; \li+rule+s and \li+definition+s remain, via a unified \li+def+
declaration devoid of any syntactic sugar. Perhaps more importantly, definitions are
provided in-order and our usage of default calculus becomes clear.

\newcommand{\codevar}[1]{\ensuremath{\mathtt{#1}}}
\newcommand{\synvar}[1]{\ensuremath{#1}}
\newcommand{\synkeyword}[1]{\textcolor{red!60!black}{\texttt{#1}}}
\newcommand{\synpunct}[1]{\textcolor{black!40!white}{\texttt{#1}}}
\newcommand{\synname}[1]{\ensuremath{\mathsf{#1}}}
\newcommand{\synbool}{\synkeyword{bool}}
\newcommand{\synnum}{\synkeyword{num}}
\newcommand{\syndate}{\synkeyword{date}}
\newcommand{\synvec}{\synkeyword{vec~}}
\newcommand{\synopt}{\synkeyword{opt~}}
\newcommand{\synrule}{\synkeyword{def\xspace}}
\newcommand{\synlet}{\synkeyword{let~}}
\newcommand{\synin}{\synkeyword{~in~}}
\newcommand{\synif}{\synkeyword{if~}}
\newcommand{\synthen}{\synkeyword{~then~}}
\newcommand{\synelse}{\synkeyword{~else~}}
\newcommand{\syncall}{\synkeyword{call~}}
\newcommand{\synscope}{\synkeyword{scope~}}
\newcommand{\synequal}{\synpunct{~=~}}
\newcommand{\synjust}{~\synpunct{:\raisebox{-0.9pt}{-}}~}
\newcommand{\syntyped}{~\synpunct{:}~}
\newcommand{\syncomma}{\synpunct{,}}
\newcommand{\syndot}{\synpunct{.}~}
\newcommand{\synunit}{\synpunct{()}}
\newcommand{\synunitt}{\synkeyword{unit}}
\newcommand{\syntrue}{\synkeyword{true}\xspace}
\newcommand{\synfalse}{\synkeyword{false}\xspace}
\newcommand{\synop}{\synpunct{\odot}}
\newcommand{\synlambda}{\synpunct{$\lambda$}~}
\newcommand{\synand}{\synpunct{\wedge}}
\newcommand{\synor}{\synpunct{\vee}}
\newcommand{\synlparen}{\synpunct{(}}
\newcommand{\synrparen}{\synpunct{)}}
\newcommand{\synlsquare}{\synpunct{[}}
\newcommand{\synrsquare}{\synpunct{]}}
\newcommand{\synlbracket}{\synpunct{\{}}
\newcommand{\synrbracket}{\synpunct{\}}}
\newcommand{\synlangle}{\synpunct{$\langle$}}
\newcommand{\synrangle}{\synpunct{$\rangle$}}
\newcommand{\synmid}{\synpunct{~$|$~}}
\newcommand{\synemptydefault}{\synvar{\varnothing}\xspace}
\newcommand{\synerror}{\synvar{\circledast}\xspace}
\newcommand{\synstar}{\synpunct{~$*$~}}
\newcommand{\synvardef}{\synkeyword{definition~}}
\newcommand{\synscopecall}{\synkeyword{scope\_call~}}
\newcommand{\synlarrow}{~\synpunct{$\leftarrow$}~}
\newcommand{\synarrow}{~\synpunct{$\rightarrow$}~}
\newcommand{\synellipsis}{\synpunct{,$\ldots$,}}
\newcommand{\synlistellipsis}{\synpunct{;$\ldots$;}}
\newcommand{\syndef}{$ ::= $}
\newcommand{\synalt}{\;$|$\;}
\newcommand{\synhole}{\synvar{\cdot}}
\newcommand{\syncrashifempty}{\synkeyword{crash\_if\_empty}}
\newcommand{\synnone}{\texttt{None}}
\newcommand{\synsome}{\texttt{Some}~}
\newcommand{\synmatch}{\synkeyword{match}~}
\newcommand{\synwith}{~\synkeyword{with}~}
\newcommand{\synoption}{\texttt{option}\;}
\newcommand{\synraise}{\synkeyword{raise}\;}
\newcommand{\synfoldleft}{\texttt{fold\_left}~}
\newcommand{\syntry}{\synkeyword{try}\;}
\newcommand{\synlist}{\texttt{list}\;}

\newcommand{\typctx}[1]{\textcolor{orange!90!black}{\ensuremath{#1}}}
\newcommand{\typempty}{\typctx{\varnothing}}
\newcommand{\typcomma}{\typctx{,\;}}
\newcommand{\typvdash}{\typctx{\;\vdash\;}}
\newcommand{\typcolon}{\typctx{\;:\;}}
\newcommand{\typlpar}{\typctx{(}}
\newcommand{\typrpar}{\typctx{)}}

\newcommand{\exctx}[1]{\textcolor{blue!80!black}{\ensuremath{#1}}}
\newcommand{\exeemptysubdefaults}{\exctx{\mathsf{empty\_count}}}
\newcommand{\execonflictsubdefaults}{\exctx{\mathsf{conflict\_count}}}
\newcommand{\Omegaarg}{\Omega_{arg}}
\newcommand{\excaller}{\exctx{\complement}}
\newcommand{\excomma}{\exctx{,}\;}
\newcommand{\exvdash}{\;\exctx{\vdash}\;}
\newcommand{\exempty}{\exctx{\varnothing}}
\newcommand{\exemptyv}{\exctx{\varnothing_v}}
\newcommand{\exemptyarg}{\exctx{\varnothing_{arg}}}
\newcommand{\exvarmap}{\exctx{~\mapsto~}}
\newcommand{\exscopemap}{\exctx{~\rightarrowtail~}}
\newcommand{\exArrow}{\exctx{~\Rrightarrow~}}
\newcommand{\exeq}{\exctx{\;=\;}}
\newcommand{\exeval}{\exctx{\;\longrightarrow\;}}
\newcommand{\exevalstar}{\exctx{\;\longrightarrow^*\;}}
\newcommand{\exat}{\exctx{\texttt{\;@\;}}}
\newcommand{\exsemicolon}{\exctx{;~}}
\newcommand{\excomp}{\dashrightarrow}

\newcommand{\redctx}[1]{\textcolor{green!50!black}{\ensuremath{#1}}}
\newcommand{\reduces}{\redctx{~\rightsquigarrow~}}
\newcommand{\redvdash}{\redctx{\;\vdash\;}}
\newcommand{\redturnstile}[1]{\;\ensuremath{\redctx{\vdash}_{#1}}\;\;}
\newcommand{\redcomma}{\redctx{,\;}}
\newcommand{\redsc}{\redctx{;\;}}
\newcommand{\redcolon}{\redctx{\;:\;}}
\newcommand{\redempty}{\redctx{\varnothing}}
\newcommand{\redproduce}{\;\redctx{\Rrightarrow}\;}
\newcommand{\redellipsis}{\redctx{,\ldots,~}}

\newcommand{\redlparen}{\redctx{(}}
\newcommand{\redrparen}{\redctx{)}}
\newcommand{\redequal}{\redctx{~=~}}
\newcommand{\redinit}{\redctx{\mathsf{init\_subvars}}}

\newcommand{\compctx}[1]{\textcolor{yellow!70!black}{\ensuremath{#1}}}
\newcommand{\compkeyword}[1]{\textcolor{yellow!60!black}{\texttt{#1}}}
\newcommand{\compiles}{\ensuremath{~\compctx{\rightrightarrows}~}}
\newcommand{\compnormal}{\compkeyword{normal}}
\newcommand{\compdefault}{\compkeyword{default}}
\newcommand{\compcons}{\compkeyword{cons}}
\newcommand{\compvdash}{\compctx{\;\vdash\;}}
\newcommand{\compok}{\;\;\compkeyword{ok}}

\begin{figure}
  \centering

\begin{center}
  \smaller
\begin{tabular}{llrll}
  Scope name&\synvar{S}\\
  \\[-2.5ex]
  Sub-scope instance&\synvar{S_n}\\
  \\[-2.5ex]
  Location&\synvar{\ell}&\syndef&\synvar{x}&scope variable\\
        &&\synalt&$\synvar{S}_\synvar{n}$\synlsquare\synvar{x}\synrsquare&sub-scope variable\\
  \\[-2.5ex]
  Type&\synvar{\tau}&\syndef&\synbool\synalt\synunitt&base types\\
  &&\synalt&\synvar{\tau}\synarrow\synvar{\tau}&function type \\
  \\[-2.5ex]
  Expression&\synvar{e}&\syndef&\synvar{x}\synalt\syntrue\synalt\synfalse\synalt\synunit&variable, literals\\
  &&\synalt&\synlambda\synlparen\synvar{x}\syntyped\synvar{\tau}\synrparen\syndot\synvar{e}\synalt\synvar{e}\;\synvar{e}&$\lambda$-calculus\\
  &&\synalt&\synvar\ell&location\\
  &&\synalt&\synvar{d}&default term\\
  \\[-2.5ex]
  Default&\synvar{d}&\syndef&\synlangle $\vec{\synvar{e}} \synmid\synvar{e}\synjust\synvar{e}$\synrangle&default term\\
  &&\synalt&\synerror&conflict error term\\
  &&\synalt&\synemptydefault&empty error term\\
  \\[-2.5ex]
  Atom&\synvar{a}&\syndef&\synrule\ \synvar{\ell}\syntyped\synvar{\tau}\synequal\synlangle
                        $\vec{\synvar{e}}$\synmid\synvar{e}\synjust
                         \synvar{e}\synrangle
      &location definition\\
  &&\synalt&\syncall$\synvar{S}_\synvar{n}$&sub-scope call\\
  \\[-2.5ex]
  Scope declaration&\synvar{\sigma}&\syndef&\synscope\synvar{S}\syntyped $\vec{\synvar{a}}$\\
  \\[-2.5ex]
  Program&\synvar{P}&\syndef&$\vec{\sigma}$\\
\end{tabular}
\end{center}

\caption{The scope language, our first intermediate representation}
  \label{fig:scope}
\end{figure}

\newcommand\synsub[2]{#1\synlsquare\synvar#2\synrsquare}
\newcommand\syndefault[3]{\synlangle $\synvar{#1} \synmid\synvar{#2}\synjust\synvar{#3}$\synrangle}

\fref{scope} presents the syntax of the scope language.
We focus on the essence of \catala, i.e. how to formalize a language with
default calculus at its core; to that end, and from this section onwards, we
omit auxiliary features, such as data types, in order to focus on a core
calculus.

To avoid carrying an environment, a reference to a sub-scope variable, such as
\li+person1.personal+ earlier, is modeled as a reference to a sub-scope
annotated with a unique identifier, such as
\li+Section121SinglePerson+$_1.\mathsf{personal}$. Therefore, locations are
either a local variable $x$, or a sub-scope variable, of the form $\synvar{S}_\synvar{n}$\synlsquare\synvar{x}\synrsquare.
Note that sub-scoping enables scope calls nesting in all generality. However,
we do not allow in our syntax references to sub-scopes' sub-scopes like
$\synvar{S}_{\synvar{n}}\synlsquare\synvar{S'}_{\synvar{n'}}\synlsquare\synvar{x}\synrsquare\synrsquare$,
as this would unnecessarily complicate our semantics model.

Types and expressions are standard, save for default terms $d$ of
the form \syndefault{\vec e_i}{e'}{e''}. This form resembles default logic terms
 $a:\vec{b_i}/c$ introduced earlier (\sref{insights}); the \emph{exceptions}
$\vec{b_i}$ become $\vec{e_i}$; the \emph{precondition} $a$ becomes $e'$; and the
\emph{consequence} $c$ becomes $e''$.
We adopt the following reduction semantics for $d$.
Each of the
exceptions $e_i$ is evaluated; if two or more are valid (i.e. not of the
form \synemptydefault), a conflict error \synerror is raised. If exactly
one exception $e_i$ is valid, the final result is $e_i$. If no exception is
valid, and $e'$ evaluates to \syntrue the final result is $e''$. If no
exception is valid, and $e'$ evaluates to \synfalse, the final result is
\synemptydefault. We provide a full formal semantics of default terms in
\S\ref{sec:defaultcalculus}.

The syntactic form \syndefault{\vec e_i}{e'}{e''} encodes a \emph{static} tree
of priorities, baking the pre-order directly in the syntax tree of each
definition. We thus offer a restricted form of prioritized default logic, in which
each definition is its own world, equipped with a static pre-order.

Atoms $a$ either define a new location, gathering all default cases and
exceptions in a single place; or, rules indicate that a sub-scope needs to
be called to compute further definitions.

We now explain how to desugar the surface syntax, presented in \sref{tutorial}, to
this scope language.

\paragraph{Syntactic sugar}

\tref{desugaring} presents rewriting rules, whose transitive closure forms our
desugaring. These rules operate within the surface language;
\tref{desugaring} abbreviates surface-level keywords for easier typesetting.

In its full generality, \catala allows exceptions to definitions, followed by an
arbitrary nesting of exceptions to exceptions. This is achieved by a label
mechanism: all exceptions and definitions are \emph{labeled}, and each
exception refers to the definition or exception it overrides. Exceptions to
exceptions are actually found in the law, and while we spared the reader in our
earlier tutorial, we have found actual use-cases where this complex scenario was
needed.
Exceptions to exceptions remain rare; the goal of our syntactic sugar is to
allow for a more compact notation in common cases, which later gets translated
to a series of fully labeled definitions and exceptions.

After
desugaring, definitions and exceptions form a forest, with exactly one root
\li+definition+s node for each variable $X$, holding an $n$-ary tree of \li+exception+
nodes.

We start with the desugaring of \li+rule+ which,
as mentioned earlier, is a boolean
definition with a base case of \li+false+ (i). Definitions without conditions
desugar to the trivial \syntrue condition (ii).

The formulation of (iiia) allows the user to provide multiple definitions for the
same variable $X$ without labeling any of them; thanks to (iiia), these are
implicitly understood to be a series of exceptions without a default case.
The surface syntax always requires a default to be provided; internally, the
\li+nodefault+ simply becomes \li+condition true consequence equals +\synemptydefault.

We provide another alternative to the fully labeled form via (iiib); the rule
allows the user to provide a single base definition, which may then be overridden via
a series of exceptions. To that end, we introduce a unique label $L_X$ which
un-annotated exceptions are understood to refer to (iv).

\begin{table}
  \centering
  \footnotesize
  \ttfamily
  \begin{tabular}{lp{.36\columnwidth}p{.52\columnwidth}}
  \toprule
  &\small\sffamily\bfseries Syntactic sugar & \small\sffamily\bfseries \ldots rewrites to \\\midrule

\sffamily(i) & rule X under cond. Y cons. fulfilled &
label L$_X$ def. X equals false \sffamily \hfill (inserted once) \newline
\ttfamily
exception L$_X$ def. X under cond. Y cons. equals true \\
\midrule

\sffamily(ii) & def. X equals Y &
def. X under cond. true cons. equals Y \\
\midrule

\sffamily (iiia) & def. X \dots \newline
\sffamily \hfill (multiple definitions of \texttt{X}, no exceptions) &
label L$_X$
def. X nodefault   \sffamily \hfill (inserted once) \newline
\ttfamily
exception L$_X$ def. X \dots\\
\midrule

\sffamily (iiib) & def. X \dots \newline
\sffamily \hfill (single definition of \texttt{X}) &
label L$_X$ def. X \dots \\
\midrule

\sffamily (iv) & exception def. X &
exception L$_X$
def. X \\

\bottomrule

  \end{tabular}
  \caption{Desugaring the surface language into an explicit surface syntax}
  \label{table:desugaring}
\end{table}

\paragraph{Materializing the default tree}

Equipped with our default expressions $d$, we show how to translate a
scattered series of \catala definitions into a single \li+def+ rule from the scope
language. We write $X,L \rightsquigarrow d$, meaning that the definition of $X$
labeled $L$, along with all the (transitive) exceptions to $L$, collectively translate to $d$. We
use an auxiliary helper $\mathsf{lookup}(X,L) = C,D,\vec L_i$,
meaning that at label $L$, under condition $C$, $X$ is defined to be $D$,
subject to a series of exceptions labeled $L_i$.

\begin{figure}
  \smaller
  \centering
  \begin{mathpar}
    \inferrule[D-Label]{
      \mathsf{lookup}(X,L) = C, D, \vec L_i \\
      X,L_i \rightsquigarrow d_i
    }{
    X, L \rightsquigarrow \syndefault{\vec d_i}CD
    }

    \inferrule[D-EntryPoint]{
      X, L \rightsquigarrow \syndefault{\vec d_i}CD
    }{
    \mathsf{label} \ L \ \mathsf{definition} \ X \ \ldots
    \rightsquigarrow \synrule\ X = \syndefault{\vec d_i}CD
    }
  \end{mathpar}
  \caption{Building the default tree and translating surface definitions}
  \label{fig:defaults}
\end{figure}

Rule \TirName{D-Label} performs the bulk of the work, and gathers the exception
labels $L_i$; each of them translates to a default expression $d_i$, all of
which appear on the left-hand side of the resulting translation; if all of the
$d_i$ are empty, the expression evaluates to $D$
guarded under condition $C$. As an illustration, if no exceptions are to be
found, the translation is simply \syndefault{}CD. Finally, rule
\TirName{D-EntryPoint} states that the translation starts at the root
\li+definition+ nodes.

\paragraph{Reordering definitions}

Our final steps consists in dealing with the fact that \synrule{}s remain
unordered. To that end, we perform two topological sorts. First, for each scope
$S$, we collect all definitions and re-order them according to a local
dependency relation $\to$:

\[
\begin{cases}
  y \to x & \text{ if } \mathsf{def}\;x = \dots y \dots \\
  S_n \to x & \text{ if } \mathsf{def}\;x = \dots S_n[y] \dots \\
  y \to S_n & \text{ if } \mathsf{def}\;S_n[x] = \dots y \dots
\end{cases}
\]

After re-ordering, we obtain a scope $S$ where definitions can be processed
linearly. Sub-scope nodes of the form $S_n$ become \synkeyword{call}s, to
indicate the position at which the sub-scope computation can be performed, i.e.
once its parameters have been filled and before its outputs are needed.

We then topologically sort the scopes themselves to obtain a linearized order.
We thus move from a declarative language to a functional language where
programs can be processed in evaluation order. In both cases, we detect the
presence of cycles, and error out. General recursion is not found in the law,
and is likely to indicate an error in modeling. Bounded recursion, which we saw
in \sref{121b}, can be manually unrolled to make it apparent.

\subsection{From the Scope Language to a Default Calculus}
\label{sec:defaultcalculus}

\begin{figure}
  \centering
  \smaller
\begin{tabular}{lrrll}
  Type&\synvar{\tau}&\syndef&\synbool\synalt\synunitt&boolean and unit types\\
  &&\synalt&\synvar{\tau}\synarrow\synvar{\tau}&function type \\
  \\[-2.5ex]
  Expression&\synvar{e}&\syndef&\synvar{x}\synalt\synvar s\synalt\syntrue\synalt\synfalse\synalt\synunit&
  variable, top-level name, literals\\
  &&\synalt&\synlambda\synlparen\synvar{x}\syntyped\synvar{\tau}\synrparen\syndot\synvar{e}\synalt\synvar{e}\;\synvar{e}&$\lambda$-calculus\\
  &&\synalt&\synvar{d}&default term\\
  \\[-2.5ex]
  Default&\synvar{d}&\syndef&\synlangle $\vec{\synvar{e}} \synmid\synvar{e}\synjust\synvar{e}$\synrangle&default term\\
  &&\synalt&\synerror&conflict error term\\
  &&\synalt&\synemptydefault&empty error term\\
  \\[-2.5ex]
  Top-level declaration&\synvar\sigma&\syndef&\synlet\synvar{s} = \synvar e\\
  \\[-2.5ex]
  Program&\synvar{P}&\syndef&$\vec{\sigma}$\\
\end{tabular}

\caption{The default calculus, our second intermediate representation}
  \label{fig:default}
\end{figure}

\begin{figure}
  \centering
  \smaller
\begin{mathpar}
  \inferrule[ConflictError]{}{\typctx{\Gamma}\typvdash\synerror\typcolon\synvar{\tau}}

  \inferrule[EmptyError]{}{\typctx{\Gamma}\typvdash\synemptydefault\typcolon\synvar{\tau}}

  \inferrule[T-Default]
  {
    \typctx{\Gamma}\typvdash\synvar{e_i}\typcolon{\tau}\\
    \typctx{\Gamma}\typvdash\synvar{e_{\text{just}}}\typcolon\synbool\\
    \typctx{\Gamma}\typvdash\synvar{e_{\text{cons}}}\typcolon\synvar{\tau}
  }
  {\typctx{\Gamma}\typvdash\synlangle
  \synvar{e_1}\synellipsis\synvar{e_n}\synmid
  \synvar{e_{\text{just}}}\synjust\synvar{e_{\text{cons}}}\synrangle\typcolon\synvar{\tau}}
\end{mathpar}

  \caption{Typing rules for the default calculus}
  \label{fig:default-typing}
\end{figure}

For the next step of our translation, we remove the scope mechanism, replacing
\synrule{}s and \synkeyword{call}s with regular $\lambda$-abstractions and
applications. The resulting language, a core lambda calculus equipped only with
default terms, is the \emph{default calculus} (\fref{default}). The typing rules
of the default calculus are standard (\fref{default-typing}); we note that the
error terms from the default calculus are polymorphic.

\paragraph{Reduction rules}

\begin{figure}
  \centering
  \smaller
  \begin{tabular}{lrrll}
    Values&\synvar{v}&\syndef&\synlambda\synlparen\synvar{x}\syntyped\synvar{\tau}\synrparen\syndot\synvar{e}&functions\\
                    &&\synalt&\syntrue\synalt\synfalse & booleans\\
                    &&\synalt&\synerror\synalt\synemptydefault&errors\\
    Evaluation &\synvar{C_\lambda}&\syndef&\synhole\;\synvar{e}\synalt\synvar{v}\;\synhole&function application evaluation\\
    contexts&&\synalt&\synlangle$\synvar{\vec v}$\synmid\synhole\synjust\synvar{e}\synrangle&default justification evaluation\\
    &&\synalt&\synlangle$\synvar{\vec v}$\synmid\syntrue\synjust\synhole \synrangle&default consequence evaluation\\
    &\synvar{C}&\syndef&\synvar{C_\lambda}&regular contexts\\
    &&\synalt&\synlangle$\synvar{\vec v}$\syncomma\synhole\syncomma$\synvar{\vec e}$\synmid
       \synvar{e}\synjust\synvar{e}\synrangle&default exceptions evaluation\\
  \end{tabular}
  \caption{Evaluation contexts for the default calculus}
  \label{fig:default-context}
\end{figure}

\begin{figure}
  \centering
  \smaller
\begin{mathpar}
   \inferrule[D-Context]
  {\synvar{e}\exeval\synvar{e'}\\ e'\notin\{\synerror,\synemptydefault\}}
  {\synvar{C}[\synvar{e}]\exeval\synvar{C}[\synvar{e'}]}

 \inferrule[D-Beta]{}{
   (\synlambda\synlparen\synvar{x}\syntyped\synvar{\tau}\synrparen\syndot{e})\;\synvar{v}
    \exeval\synvar{e}[\synvar{x}\mapsto\synvar{v}]
 }

\inferrule[D-ContextConflictError]
  {\synvar{e}\exeval\synerror}
  {\synvar{C}[\synvar{e}]\exeval\synerror}

   \inferrule[D-ContextEmptyError]
  {\synvar{e}\exeval\synemptydefault}
  {\synvar{C_\lambda}[\synvar{e}]\exeval\synemptydefault}

  \inferrule[D-DefaultTrueNoExceptions]
  {}
  {\synlangle \synemptydefault{}\synellipsis\synemptydefault{}\synmid\syntrue\synjust \synvar{v}\synrangle\exeval v}

  \inferrule[D-DefaultFalseNoExceptions]
  {}
  {\synlangle \synemptydefault{}\synellipsis\synemptydefault{}\synmid\synfalse\synjust \synvar{e} \synrangle\exeval \synemptydefault{}}

  \inferrule[D-DefaultOneException]
  {v \neq \synemptydefault}
  {\synlangle \synemptydefault\synellipsis\synemptydefault\syncomma\synvar{v}\syncomma\synemptydefault\synellipsis\synemptydefault
  \synmid  \synvar{e_1}\synjust \synvar{e_2}
  \synrangle\exeval \synvar{v}}

  \inferrule[D-DefaultExceptionsConflict]
  {v_i \neq \synemptydefault \\ v_j \neq \synemptydefault}
  {\synlangle \synpunct{$\ldots$,}\synvar{v_i}\synellipsis\synvar{v_j}\synpunct{,$\ldots$}\synmid
  \synvar{e_1}\synjust \synvar{e_2}\synrangle\exeval \synerror{}}
\end{mathpar}

  \caption{Reduction rules for the default calculus}
  \label{fig:default-reduction}
\end{figure}

We present small-step operational semantics, of the form
\fbox{\synvar{e}\exeval\synvar{e'}}. For efficiency, we describe reduction under
a context, using a standard notion of value (\fref{default-context}), which
includes our two types of errors, \synerror and \synemptydefault. We
intentionally distinguish regular contexts $C_\lambda$ from general contexts
$C$.

\fref{default-reduction} presents the reduction rules for the default calculus.
Rule \TirName{D-Context} follows standard call-by-value reduction rules for
non-error terms; \TirName{D-Beta} needs no further comment. \synerror is made fatal by
\TirName{D-ContextConflictError}: the reduction aborts,
under \emph{any context} $C$. The behavior of \synemptydefault~is different:
such an error propagates only up to its enclosing “regular” context $C_\lambda$;
this means that such an \synemptydefault-error can be caught, as long as it
appears in the exception list of an enclosing default expression. Therefore, we
now turn our attention to the rules that govern the evaluation of default
expressions.

If no exception is valid, i.e. if the left-hand side contains only
\synemptydefault{}s; and if after further evaluation, the justification is \syntrue for the consequence $v$,
then the whole default reduces to $v$ (\TirName{D-DefaultTrueNoExceptions}). If
no exception is valid, and if the justification is \synfalse, then we do not
need to evaluate the consequence, and the default
is empty, i.e. the expression reduces to \synemptydefault. If \emph{exactly} one
exception is a non-empty value $v$, then the default reduces to $v$. In that
case, we evaluate neither the justification or the consequence
(\TirName{D-DefaultOneException}). Finally, if two or more exceptions are
non-empty, we cannot determine the priority order between them, and abort
program execution (\TirName{D-DefaultExceptionsConflict}).

\paragraph{Compiling the scope language}

\begin{figure}
  \centering
  \smaller
  \begin{mathpar}
    \inferrule[C-Scope]{
      \mathsf{local\_vars}(S) = \overrightarrow{x: \tau} \\
      \mathsf{calls}(S) = \vec S \\\\
      \mathsf{local\_vars}(S_i) = \overrightarrow{S_i[x]} \\
      S, [] \vdash \vec a \hookrightarrow e
    }{
    \begin{array}l
    \synscope S : \vec a \hfill \hookrightarrow \cr
    \synlet S (\overrightarrow{x: \synunit \to\tau}) = %
    \; \overrightarrow{\synlet (\overrightarrow{S_i[x]}) =
    (\overrightarrow{\synlambda\synunit\syndot\synemptydefault}) \synin}%
    \; e
    \end{array}
    }

    \inferrule[C-Empty]{
      \mathsf{local\_vars}(S) = \vec x
    }{
    S, \Delta \vdash [] \hookrightarrow (\overrightarrow{\mathsf{force}(\Delta,x)})
    }

    \inferrule[C-Def]{
    S, \ell \cdot \Delta \vdash \vec a \hookrightarrow e_{\mathsf{rest}}
    }{
    \begin{array}l
      S, \Delta \vdash \synrule~\ell = e :: \vec a \hfill \hookrightarrow \cr
      \synlet \ell = \syndefault{\ell\ \synunit}\syntrue e \synin %
      e_{\mathsf{rest}}
    \end{array}
    }

    \inferrule[C-Call]{
      S \neq S_i \\
      S, \overrightarrow{S_i[x]} \cdot \Delta \vdash \vec a \hookrightarrow e_{\mathsf{rest}} \\\\
      \mathsf{local\_vars}(S_i) = \overrightarrow{S_i[x]}
    }{
    \begin{array}l
      S, \Delta \vdash \syncall S_i :: \vec a \hfill \hookrightarrow \cr
      \synlet (\overrightarrow{S_i[x]}) = S_i (\overrightarrow{\mathsf{thunk}(\Delta,S_i[x])}) \synin %
      e_{\mathsf{rest}}
    \end{array}
    }

    \inferrule[F-In]{
      x \in \Delta
    }{
      \mathsf{force}(\Delta,x) = x
    }

    \inferrule[F-NotIn]{
      x \not\in \Delta
    }{
      \mathsf{force}(\Delta,x) = x\;\synunit
    }

    \inferrule[T-In]{
      x \in \Delta
    }{
      \mathsf{thunk}(\Delta,x) = \synlambda\synunit\syndot x
    }

    \inferrule[T-NotIn]{
      x \not\in \Delta
    }{
      \mathsf{thunk}(\Delta,x) = x
    }
  \end{mathpar}
  \caption{Compiling the scope language to a default calculus}
  \label{fig:scope-compile}
\end{figure}

We succinctly describe the compilation of the scope language to the default
calculus in \fref{scope-compile}. Our goal is to get rid of scopes in favor of
regular lambda-abstractions, all the while preserving the evaluation semantics;
incorrect order of evaluation might lead to propagating premature errors (i.e.
throwing exceptions too early).

We assume for simplicity of presentation that we are equipped with tuples, where $(x_1, \dots,
x_n)$ is concisely written $(\vec x)$. We also assume that we are equipped with
let-bindings, of the form $\synlet (x_1, \dots, x_n) = e$, for which we adopt
the same concise notation.
Given a scope $S$ made up of atoms $\vec a$, $\mathsf{local\_vars}(S)$ returns
all variables $x$ for which $\synrule~x \in \vec a$. Note that this does not
include rules that override sub-scope variables, which are of the form
$\synrule~S_n[x]$. We now turn to our judgement, of the form \fbox{$S,\Delta\vdash a
\hookrightarrow e$}, to be read as “in the translation of scope $S$, knowing
that variables in $\Delta$ have been forced, scope rule $r$ translates to
default calculus expression $e$”.

A scope $S$ with local variables $\vec x$ compiles to a function that takes an
$n$-tuple $(\vec x)$ containing \emph{potential overrides} for all of its context
variables (\TirName{C-Scope}). In the translation, each $x$ therefore becomes a
thunk, so as to preserve reduction semantics: should the caller decide to leave
a local variable $x_i$ to be \synemptydefault, having a thunk prevents
\TirName{D-ContextEmptyError} from triggering and returning prematurely. Rule
\TirName{C-Scope} performs additional duties. For each one of the sub-scopes
$S_i$ used by $S$, we set all of the arguments to $S_i$, denoted
$\overrightarrow{S_i[x]}$, to be a thunked \synemptydefault to start with.

Advancing through the scope $S$, we may encounter definitions or calls. For
definitions (\TirName{C-Def}), we simply insert a shadowing let-binding, and
record that $\ell$ has been forced by extending $\Delta$. Whether
$\ell$ is of the form $x$ or $S_n[x]$, we know that the previous binding was
thunked, since our previous desugaring guarantees that any variable $\ell$ now
has a single definition. The rewritten default expression gives the caller-provided argument
higher precedence; barring any useful information provided by the caller, we
fall back on the existing definition $e$. This key step explains how our law-centric
syntax, which allows caller scopes to override variables local to a callee
scope, translates to the default calculus.

For calls (\TirName{C-Call}), we ensure all of the arguments are thunked before
calling the sub-scope; the return tuple contains \emph{forced} values, which we
record by extending $\Delta$ with all $\overrightarrow{S_i[x]}$. The premise $S
\neq S_i$ captures the fact that recursion is not allowed.

Finally, after all rules have been translated and we are left with nothing but
the empty list $[]$ (\TirName{C-Empty}), we simply force all scope-local
variables $\vec x$ and return them as a tuple.

\subsection{From the Default Calculus to a Lambda Calculus}
\label{sec:lambdacalculus}

While sufficient to power the \catala surface language, the
relatively simple semantics of our default calculus are non-standard. We now
wish to compile to more standard constructs found in existing programming
languages.
We remark that the reduction semantics for default terms resembles that of
exceptions: empty-default errors propagate (“are thrown”) only up to the
enclosing default term (“the try-catch”).
Confirming this intuition and providing a safe path from \catala
to existing programming languages, we now present a compilation scheme from the
default calculus to a lambda calculus enriched with a few standard additions:
lists, options and exceptions.

\begin{figure}
  \centering
  \smaller
\begin{tabular}{lrrll}
  Type&\synvar{\tau}&\syndef&\synbool\synalt\synunitt&boolean and unit types\\
  &&\synalt&\synvar{\tau}\synarrow\synvar{\tau}&function type \\
  &&\synalt&\synlist\synvar{\tau}&list type\\
  &&\synalt&\synoption\synvar{\tau}&option type\\
  \\[-2.5ex]
  Expression&\synvar{e}&\syndef&\synvar{x}\synalt\synvar s\synalt\syntrue\synalt\synfalse\synalt\synunit&
  variable, top-level name, literals\\
  &&\synalt&\synlambda\synlparen\synvar{x}\syntyped\synvar{\tau}\synrparen\syndot\synvar{e}\synalt\synvar{e}\;\synvar{e}&$\lambda$-calculus\\
  &&\synalt&\synnone\synalt\synsome\synvar{e}&option constructors\\
  &&\synalt&\synmatch\synvar{e}\synwith&option destructuring\\
  &&       &\quad\synalt\synnone\synarrow\synvar{e}\synalt\synsome\synvar{x}\synarrow\synvar{e}\\
  &&\synalt&\synlsquare$\vec{\synvar{e}}$\synrsquare\synalt\synfoldleft\synvar{e}~\synvar{e}~\synvar{e}&list introduction and fold\\
  &&\synalt&\synraise \synvar{\varepsilon}\synalt\syntry\synvar{e}\synwith\synvar{\varepsilon}\synarrow\synvar{e}&exceptions\\
  \\[-2.5ex]
  Exception&\synvar{\varepsilon}&\syndef&\synemptydefault&empty exception\\
  &&\synalt&\synerror&conflict exception\\
  \\[-2.5ex]
  Top-level declaration&\synvar\sigma&\syndef&\synlet\synvar{s} = \synvar e\\
  \\[-2.5ex]
  Program&\synvar{P}&\syndef&$\vec{\sigma}$\\
\end{tabular}

\caption{The enriched lambda calculus, our final translation target}
  \label{fig:lambdasyntax}
\end{figure}

\fref{lambdasyntax} shows the syntax of the target lambda calculus. In order to
focus on the gist of the translation, we introduce \texttt{list} and
\texttt{option} as special, built-in datatypes, rather than a full facility for
user-defined inductive types. For those reasons, we offer the minimum set of
operations we need: constructors and destructors for \texttt{option}, and a left
fold for \texttt{list}s.
We omit typing and reduction rules, which are standard.
The only source term that does not belong to the target lambda calculus is
the default term \syndefault{\vec{e}}{e_\mathrm{just}}{e_\mathrm{cons}}. Hence,
translating this term is the crux of our translation.

Our translation is of the form \fbox{\synvar{e}\compiles\synvar{e'}}, where \synvar{e}
is a term of the default calculus and \synvar{e'} is a term of the target
lambda calculus. \fref{translationrules} presents our translation scheme.
The semantics of default
terms are intertwined with those of \synemptydefault and \synerror. The
translation of \synemptydefault and \synerror is simple: both compile to
exceptions in the target language.
We now focus on \TirName{C-Default}, which deals with default terms. As a
warm-up, we start with a special case:
\syndefault{}{e_\mathrm{just}}{e_\mathrm{cons}}. We translate this term to \synif\synvar{e_\mathrm{just}}
\synthen\synvar{e_\mathrm{cons}}\synelse\synraise\synemptydefault, which obeys
the evaluation semantics of both \TirName{D-DefaultTrueNoExceptions}
and \TirName{D-DefaultFalseNoExceptions}. This simple example serves as a
blueprint for the more general case, which has to take into account the list of
exceptions $\vec e$, and specifically count how many of them are
\synemptydefault.

\begin{figure}
\smaller
\begin{mathpar}
  \inferrule[C-Default]{
    \synvar{e_1}\compiles\synvar{e_1'}\\
    \cdots\\
    \synvar{e_n}\compiles\synvar{e_n'}\\
    \synvar{e_\mathrm{just}}\compiles\synvar{e_\mathrm{just}'}\\
    \synvar{e_\mathrm{cons}}\compiles\synvar{e_\mathrm{cons}'}\\
  }{
    \synlangle\synvar{e_1}\synellipsis\synvar{e_n}\synmid\synvar{e_\mathrm{just}}
    \synjust\synvar{e_\mathrm{cons}}\synrangle\compiles\\
    \synlet\synvar{r_\mathrm{exceptions}}\synequal
    \texttt{process\_exceptions}\;\synlsquare\synlambda\synvar{\_}\synarrow\synvar{e_1'}
    \synlistellipsis\synlambda\synvar{\_}\synarrow\synvar{e_n'}\synrsquare\synin\\\synmatch
    \synvar{r_\mathrm{exceptions}}\synwith\synsome\synvar{e'}\synarrow
    \synvar{e'}\synmid\synnone\synarrow
    \synif\synvar{e_\mathrm{just}'}\synthen\synvar{e_\mathrm{cons}'}\synelse\synraise\synemptydefault
  }

  \inferrule[C-EmptyError]{}{
    \synemptydefault\compiles\synraise\synemptydefault
  }

  \inferrule[C-ConflictError]{}{
    \synerror\compiles\synraise\synerror
  }

  \inferrule[C-Var]{}{\synvar{x}\compiles\synvar{x}}

  \inferrule[C-Literal]{\synvar{e}\in\{\synunit,\;\syntrue,\;\synfalse\}}{
    \synvar{e}\compiles\synvar{e}
  }

  \inferrule[C-Abs]{
    \synvar{e}\compiles\synvar{e'}
  }{
    \synlambda\synlparen\synvar{x}\syntyped\synvar{\tau}\synrparen\syndot\synvar{e}\compiles
    \synlambda\synlparen\synvar{x}\syntyped\synvar{\tau}\synrparen\syndot\synvar{e'}
  }

  \inferrule[C-App]{
    \synvar{e_1}\compiles\synvar{e_1'}\\
    \synvar{e_2}\compiles\synvar{e_2'}
  }{
    \synvar{e_1}\;\synvar{e_2}\compiles\synvar{e_1'}\;\synvar{e_2'}
  }
  \end{mathpar}
\caption{Translation rules from default calculus to lambda calculus\label{fig:translationrules}}
\end{figure}

In the general case, \TirName{C-Default} relies on a helper,
\texttt{process\_exceptions}; each exception is translated, thunked, then passed
to the helper; if the helper returns \synsome, exactly one exception did not
evaluate to \synemptydefault; we return it. If the helper returns \synnone, no
exception applied, and we fall back to the simple case we previously described.

We now review \li+process_exceptions+ defined in \fref{processexception}.
It folds over the list of exceptions, with the accumulator initially set
to \synnone, meaning no applicable exception was found. Each exception is
forced in order, thus implementing the reduction semantics of the default
calculus. The accumulator transitions from \synnone{} to \synsome if a
non-empty exception is found, thus implementing a simple automaton that counts
the number of non-empty exceptions. If two non-\synemptydefault exceptions are found, the
automaton detects an invalid transition and aborts with a non-catchable \synerror.

\begin{figure}
  \smaller
\begin{align*}
  \texttt{process\_exceptions}\quad&\syntyped&&
  \synlist\synlparen\synunitt\synarrow\synvar{\tau}\synrparen\synarrow\synoption\synvar{\tau}\\
  \texttt{process\_exceptions}\quad&\triangleq&&
  \texttt{fold\_left}\;\synlparen\synlambda\synlparen\synvar{a}\syntyped
  \synoption\synvar{\tau}\synrparen\;
  \synlparen\synvar{e'}\syntyped
  \synunitt\synarrow\synvar{\tau}\synrparen\syndot\\
  &&&\quad\synlet\synvar{e'}\syntyped\synvar{\tau}\synequal\syntry
  \synsome\synlparen\synvar{e'}\synunit\synrparen
  \synwith\synemptydefault\synarrow \synnone\synin\\
  &&&\quad\synmatch\synlparen\synvar{a}\syncomma\;\synvar{e'}\synrparen\synwith\\
  &&&\quad\quad\synmid\synlparen\synnone\syncomma\;\synvar{e'}\synrparen\synarrow\synvar{e'}\\
  &&&\quad\quad\synmid\synlparen\synsome\synvar{a}\syncomma\;\synnone\synrparen\synarrow
  \synsome\synvar{a}\\
  &&&\quad\quad\synmid\synlparen\synsome\synvar{a}\syncomma\;\synsome
  \synvar{e'}\synrparen\synarrow \synraise\synerror \synrparen\;\synnone
\end{align*}
\caption{\texttt{process\_exceptions} translation helper\label{fig:processexception}}
\end{figure}

\subsection{Discussion of Design Choices}
The current shape of \catala represents the culmination of a long design process
for the language. We now discuss a few of the insights we gained as we iterated
through previous versions of \catala.

For the surface language, a guiding design principle was to always guarantee
that the \catala source code matches the structure of the law. We have managed
to establish this property not only for Section 121~(\sref{tutorial}), but also
for all of the other examples we currently have in the \catala repository.
To achieve this, a key insight was the realization that every piece of statutory
law we looked at (in the US and French legal systems) follows a general case /
exceptions drafting style. This style, which we saw earlier (\sref{tutorial}),
means that the law encodes a \emph{static} priority structure. Intuitively,
computing a given value boils down to evaluating an n-ary \emph{tree} of
definitions, where nodes and edges are statically known. The children of a given
node are mutually-exclusive \emph{exceptions} to their parent definition; either
one exception applies, and the computation stops. Or if no exception applies,
the parent definition is evaluated instead. This recursive evaluation proceeds
all the way up to the root of the tree, which represents the initial default
definition.

The surface language was crafted to support encoding that tree of exceptions
within the syntax of the language via the label mechanism. This informed our
various choices for the syntactic sugar; notably, we make it easy to define $n$
mutually-exclusive definitions in one go thanks to syntactic sugars (i) and
(iiia) in Table~\ref{table:desugaring}.

A consequence of designing \catala around a \emph{static} tree of exceptions for
each defined value is that the internal default calculus representation was
\emph{drastically} simplified. In the original presentation of prioritized default logic,
values have the form $\langle e_1, \dots, e_n | e_c : e_d |\leqslant \rangle$ where
$\leqslant$ is a pre-order that compares the $e_i$ \emph{at run-time} to
determine the order of priorities. We initially adopted this very general
presentation for \catala, but found out that this made the semantics nearly
impossible to explain; made the rules overly complicated and the proof of
soundness very challenging; and more importantly, was not necessary to capture
the essence of Western statutory law. Dropping a run-time pre-order $\leqslant$
was the key insight that made the internal default calculus representation
tractable, both in the paper formalization and in the formal proof.

The scope language was introduced to eliminate the curious scoping rules and
parent-overrides of definitions, which are unusual in a lambda-calculus. We
initially envisioned a general override mechanism that would allow a parent
scope to override a sub-scope definition at any depth; that is, not just
re-define sub-scope variables, but also sub-sub-scope variables and so on. We
were initially tempted to go for this most general solution; however, we have yet
to find a single example of statutory law that needs this feature; and allowing
sub-sub-scope override would have greatly complicated the translation step. We
eventually decided to not implement this feature, to keep the compiler, the
paper rules and the formalization simple enough.

All of those insights stemmed from a close collaboration with lawyers, and the
authors of this paper are very much indebted to Sarah Lawsky and Liane Huttner
for their invaluable legal insights. Barring any legal expertise, we would have
lacked the experimental evaluation that was able to justify our simplification
choices and revisions of the language.

\subsection{Certifying the Translation}

The
translation from scope language to default calculus focuses on
turning scopes into the lambda-abstractions that they truly are underneath the
concrete syntax. This is a mundane transformation, concerned mostly
with syntax. The final step from default calculus to lambda calculus
with exceptions is much more delicate, as it involves compiling custom
evaluation semantics. To rule out any errors in the most sensitive compilation
step of \catala, we formally prove our translation correct, using
\fstar~\cite{mumon,metafstar,ahman2017dijkstra}, a proof assistant based on
dependent types, featuring support for semi-automated reasoning via the
SMT-solver Z3~\cite{z3}.

\paragraph{Correctness statement}

We wish to state two theorems about our
translation scheme. First, typing is preserved: if \codevar{de}\compiles\codevar{le}
and \typempty\typvdash\codevar{de}\typcolon\codevar{dtau}, then
\typempty\typvdash\codevar{le}\typcolon\codevar{ltau} in the target lambda calculus
where \codevar{ltau} is the (identity) translation of \codevar{dtau}.
Second, we want to establish a simulation result, i.e. the compiled program
\codevar{le} faithfully simulates a reduction step from the source language,
using $n$ steps in the target language.%

\begin{figure}
  \smaller
  \hspace*{\fill}
\begin{tikzpicture}[node distance=1.9cm]
\node (TL) {\codevar{de}};
\node[right of=TL] (TR) {\codevar{de'}};
\node[below of=TL] (BL) {\codevar{le}};
\node[right of=BL] (BR) {\codevar{le'}};
\draw (TL) edge[draw=none] node {\exeval} (TR);
\draw (TL) edge[draw=none] node[rotate=-90] {\compiles} (BL);
\draw (BL) edge[draw=none] node {$\exeval^{\!\!\!*}$} (BR);
\draw (TR) edge[draw=none] node[rotate=-90] {\compiles} (BR);
\draw (TL) edge[draw=none] node {A} (BR);
\end{tikzpicture}
\hspace*{\fill}
\begin{tikzpicture}[node distance=1cm]
  \node (TL) {\codevar{de}};
  \node[right of=TL] (TR) {\codevar{de'}};
  \node[below of=TL] (BL) {\codevar{le}};
  \node[right of=BL] (BR) {\codevar{le'}};
  \node[below of=BR] (BBR) {\codevar{target}};
  \draw (TL) edge[draw=none] node {\exeval} (TR);
  \draw (TL) edge[draw=none] node[rotate=-90] {\compiles} (BL);
  \draw (TR) edge[draw=none] node[rotate=-90] {\compiles} (BR);
  \draw (BR) edge[draw=none] node[rotate=-90] {$\exeval^{\!\!\!*}$} (BBR);
  \draw (BL) edge[draw=none] node[rotate=-45] {$\exeval^{\!\!\!*}$} (BBR);
  \draw (TL) edge[draw=none] node {B} (BR);
\end{tikzpicture}
  \hspace*{\fill}
  \caption{Translation correctness theorems\label{fig:correctnesstheorem}. A
  shows a regular simulation; B shows our variant of the theorem.}
\end{figure}
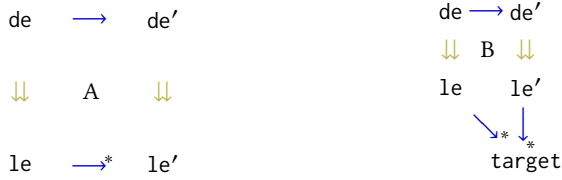

The usual simulation result is shown in \fref{correctnesstheorem}, A.
If \codevar{de} is a term of the default calculus
and if \codevar{de}\exeval\codevar{de'}, and
\codevar{de}\compiles\codevar{le}, then there exists a term
\codevar{le'} of the lambda calculus such that
\codevar{le}$\exeval^*$\codevar{le'} and \codevar{de'}\compiles\codevar{le'}.
This specific theorem does not apply in our case, because of the thunking we
introduce in our translation. As a counter-example, consider
the reduction of \synvar{e_1} within default term
\syndefault{\synvar{v_0}\syncomma\synvar{e_1}}{\synvar{e_\mathrm{just}}}{\synvar{e_\mathrm{cons}}}.
If \synvar{e_1} steps to \synvar{e_1'} in the default calculus, then the whole term
steps to \syndefault{\synvar{v_0}\syncomma\synvar{e_1'}}{\synvar{e_\mathrm{just}}}{\synvar{e_\mathrm{cons}}}.
However, we translate exceptions to thunks; and our target lambda calculus does
not support strong reduction, meaning
\synlambda\synvar{\_}\synarrow\synvar{e_{\lambda,1}}
does not step into \synlambda\synvar{\_}\synarrow\synvar{e'_{\lambda,1}}.
Diagram A is therefore not applicable in our case.

The theorem that actually holds in our case is shown as diagram B
(\fref{correctnesstheorem}). The two translated terms \codevar{le} and
\codevar{le'} eventually reduce to a common form \codevar{target}. Taking the
transitive closure of form B, we obtain that if \codevar{de} reduces to a value \codevar{dv},
then its translation \codevar{le} reduces to a value \codevar{lv} that is the
translation of \codevar{dv}, a familiar result.

\paragraph{Overview of the proof}
We have mechanically formalized the semantics of both the default calculus and target
lambda calculus, as well as the translation scheme itself, inside the \fstar
proof assistant. \fref{theoremfstar} shows the exact theorem we prove, using
concrete \fstar syntax; the theorem as stated establishes both type preservation
and variant B, via the \li+take_l_steps+ predicate and the existentially
quantified \li+n1+ and \li+n2+. We remark that if the starting term \codevar{de} is a value to
start with, we have $\codevar{le}= \codevar{translate\_exp}\;\codevar{de}$.
Inspecting \codevar{translate\_exp} (elided), we establish that source values
translate to identical target values.

\let\FancyVerbStartString\relax
\let\FancyVerbStopString\relax

\begin{figure}
  \begin{minted}[escapeinside=!!, fontsize=\small]{OCaml}
module D = DefaultCalculus; module L = LambdaCalculus
val translation_correctness (de: D.exp) (dtau: D.ty) : Lemma
  (requires (D.typing D.empty de dtau)) (ensures (
    let le = translate_exp de in let ltau = translate_ty dtau in
    L.typing L.empty le ltau !$\wedge$! begin
      if D.is_value de then L.is_value le else begin
        D.progress de dtau; D.preservation de dtau;
        let de' = Some?.v (D.step de) in
        translation_preserves_empty_typ de dtau; translation_preserves_empty_typ de' dtau;
        let le' : typed_l_exp ltau = translate_exp de' in
        exists (n1 n2:!$\mathbb{N}$!) (target: typed_l_exp ltau).
          (take_l_steps ltau le n1 == Some target !$\wedge$!
           take_l_steps ltau le' n2 == Some target) end end))
  \end{minted}
  \caption{Translation certification theorem, in \fstar\label{fig:theoremfstar}}
  \end{figure}

\paragraph{Proof effort and engineering}

Including the proof of type safety for the source and target language, our \fstar
mechanization amounts to approximately 3,500 lines of code and required
1 person-month.
We rely on partial automation via Z3, and the total verification time for the
entire development is of the order of a few minutes.  The choice of \fstar was
not motivated by any of its advanced features, such as its effect system: the
mechanization fits inside the pure fragment of \fstar. Our main motivation was
the usage of the SMT solver which can typically perform a fair amount of
symbolic reasoning and definition unrolling, thus decreasing the amount of
manual effort involved.

To focus the proof effort on the constructs that truly matter (i.e. default
expressions), the semantics of lists, folds and options are baked into the
target calculus. That is, our target calculus does not support user-defined
algebraic data types. We believe this is not a limitation, and instead allows
the proof to focus on the key parts of the translation.
We use De Bruijn indices for our binder representation, since the unrolling of
\texttt{process\_exceptions} results in variable shadowing.
Given those simplifications, we were surprised to find that our proof
still required 3,500 lines of \fstar. A lot of the complexity budget was spent on
the deep embedding of the \texttt{process\_exceptions} helper. It is during the
mechanization effort that we found out that theorem A does not hold, and that we
need to establish B instead.
Our mechanized proof thus significantly increases our confidence in the \catala
compilation toolchain; the proof is evidence that even for a small calculus and
a simple translation, a lot of subtleties still remain.

While \fstar extracts to OCaml, we chose \emph{not} to use the extracted \fstar
code within the \catala compiler. First, the proof does not take into account
all language features. Second, the actual translation occupies about 100 lines
of code in both the production \catala compiler and the proof; we are content
with comparing both side-by-side. Third, the \catala compiler features advanced
engineering for locations, error messages, and propagating those to the proof
would be difficult.

\section{The \catala Compiler}
\label{sec:compiler}

Based on this formalization, we implement \catala in a standalone compiler
and interpreter. The architecture of the compiler is based on a series of
intermediate representations, in the tradition of CompCert~\cite{Leroy} or
Nanopass~\cite{keep2013nanopass}. \fref{architecture} provides a high-level
overview of the architecture, with links to relevant sections alongside
each intermediate representation.

The compiler is written in OCaml and features approximately 13,000 lines of
code. This implementation, available as open-source software on
\href{https://github.com/CatalaLang/catala}{GitHub} and in the artifact
accompanying this paper \cite{merigoux_denis_2021_4775161},
makes good use of the rich and state-of-the art
library ecosystem for compiler writing, including \li+ocamlgraph+~\cite{conchon2007designing}
for the e.g. the two topological sorts we saw (\sref{desugaring}),
\li+bindlib+~\cite{lepigre2018abstract} for efficient and safe manipulation
of variables and terms, and the \li+menhir+ parser
generator~\cite{menhir}. Thanks to these libraries, we estimate that the
development effort was 5 person-months.

\begin{figure}
  \centering
  \smaller
  \begin{tikzpicture}[scale=0.85]
    \draw [draw=red, fill=red!20!white] (1.25,-1.5) rectangle (10.5,0.5);

    \node[draw, fill=gray!20!white, thick] (Z) at (0,0) {Source code};

    \node[draw, fill=gray!20!white, thick] (A) at (3,0) {Surface language};

    \node[draw, fill=gray!20!white, thick] (B) at (3,-1) {Desugared language};

    \node[draw, fill=gray!20!white, thick] (C) at (6.5,-1) {Scope language};

    \node[draw, fill=gray!20!white, thick] (D) at (6.5,0) {Default calculus};

    \node[draw, fill=gray!20!white, thick] (E) at (9.5,0) {$\lambda$-calculus};

    \node[draw, fill=gray!20!white, thick] (F) at (8.5,-2) {OCaml};

    \node[draw, fill=gray!20!white, thick] (G) at (10.5,-2) {Interpeter};

    \node[draw, fill=gray!20!white, thick] (H) at (8.5,-3) {JavaScript};

    \path[draw, ultra thick, ->] (Z) -- (A);
    \path[draw, ultra thick, ->] (A) -- (B);
    \path[draw, ultra thick, ->] (B) -- (C);
    \path[draw, ultra thick, ->] (C) -- (D);
    \path[draw, ultra thick, ->] (D) -- (E);
    \path[draw, ultra thick, ->] (E) -- (F);
    \path[draw, ultra thick, ->] (E) -- (G);
    \path[draw, ultra thick, ->] (F) -- (H);

    \node at ($ (A) + (0,0.75) $) {\sref{tutorial}};
    \node at ($ (B) + (0,-0.75) $) {\sref{desugaring}};
    \node at ($ (C) + (0,-0.75) $) {\sref{desugaring}};
    \node at ($ (D) + (0,0.75) $) {\sref{defaultcalculus}};
    \node at ($ (E) + (0,0.75) $) {\sref{lambdacalculus}};
    \node at ($ (H) + (-2.8,0) $) {\emph{via} \texttt{js\_of\_ocaml} \cite{vouillon2014bytecode}};
  \end{tikzpicture}
\caption{High-level architecture of the \catala compiler (red box)\label{fig:architecture}}
\end{figure}

\subsection{Usability}

We devoted a great of attention towards the usability of the compiler.
Indeed, while we don't expect lawyers to use \catala unaccompanied, we
would not want to restrict its usage to $\lambda$-savvy functional programmers.
To improve the programmer experience, we use the special parser error reporting scheme of
\li+menhir+~\cite{pottier2016reachability}, to provide customized and context-aware
syntax error messages that depend on the set of tokens acceptable by the grammar
at the site of the erroneous token (see \fref{syntaxerror}).
The shape of the error messages is heavily inspired by the Rust compiler
design~\cite{rusterrors}. Error messages flow through the compiler code via a
unique exception containing the structured data of the error message:

\begin{figure}
\begin{Verbatim}[fontsize=\small]
[ERROR] Syntax error at token "years"
[ERROR] Message: expected a unit for this literal, or a valid operator
[ERROR]          to complete the expression
[ERROR] Autosuggestion: did you mean "year", or maybe "or", or maybe "and",
[ERROR]                 or maybe "day", or maybe ".", or maybe ">", [...]
[ERROR] Error token:
[ERROR]    --> section_121.catala_en
[ERROR]     |
[ERROR] 180 |       if date_of_sale_or_exchange <=@ period.begin +@ 5 years then
[ERROR]     |                                                         ^^^^^
[ERROR] Last good token:
[ERROR]    --> section_121.catala_en
[ERROR]     |
[ERROR] 180 |       if date_of_sale_or_exchange <=@ period.begin +@ 5 years then
[ERROR]     |                                                       ^
\end{Verbatim}
\caption{Example of \catala syntax error message\label{fig:syntaxerror}}
\end{figure}

\begin{minted}[fontsize=\small]{ocaml}
exception StructuredError of (string * (string option * Pos.t) list)
\end{minted}

\noindent
This structure enables on-the-fly rewriting of error messages as they propagate up
the call stack, which is useful for e.g. adding a new piece of context linking
to a code position of a surrounding AST node. In this spirit, error messages for scope variable cycle
detection display the precise location for where the variables
in the cycle are used; error messages for default logic conflict errors
\synerror show the location
of the multiple definitions that apply at the same time for a unique variable
definition.

Finally, we have instrumented the \catala interpreter with helpful debugging
features. Indeed, when pair programming with a lawyer and a programmer over
the formalization of a piece of law, it is helpful to see what the execution
of the program would look like on carefully crafted test cases. While test
cases can be directly written in \catala using a top-level scope that simply
defines the arguments of the sub-scope to test, the compilation chain inserts
special log hooks at critical code points. When executing a test case, the interpreter
then displays a meaningful log
featuring code positions coupled with the position inside the legal statute
for each default logic definition taken.

We believe that this latter feature can easily be extended to provide a comprehensive
and legal-oriented explanation of the result of a \catala program over a particular
input. Such an explanation would help increase trust of the system by its users,
e.g. citizens subject to a complex taxation regime; thereby constituting
a concrete instance of a much sought-after \enquote{explainable AI}~\cite{goebel2018explainable, doran2017does}.

\subsection{Performance}

We ran the French family benefits \catala
program described in \sref{familybenefits}; it is as complex as
Section 121 of the US Internal Revenue Code but featuring
approximately 1500 lines of \catala code (literate programming included).
Given the description of a French household, we benchmark the time required to
compute the amount of monthly family benefits owed.

The \catala interpreter for this program runs in approximately 150ms.
We conclude that the performance of the interpreter remains acceptable even
for a production environment. When the \catala code is compiled to OCaml,
execution time drops to 0.5ms. Therefore, we conclude that performance
problems are, at this stage of the project, nonexistent.

\subsection{Extensible Compiler Backend}

A crucial consideration in designing a DSL is the interoperability story within
existing environments. While some DSLs operate in isolation, we envision \catala
programs being exposed as reusable libraries that can be called from any
development platform, following the needs of our adopters.
In the context of legal systems, this is a very strong requirement: such
environments oftentimes include
legacy, mainframe-based systems operating in large private organizations or
government agencies~\cite{newjerseycobol}.
Furthermore, since the algorithms that \catala is designed
to model are at the very heart of e.g. tax collection systems, proposing
a classic interoperability scheme based on APIs or inter-language FFIs might
create an undesirable barrier to adoption; a system designed in the 1960s
probably has no notion of API or FFI whatsoever!

Instead, we choose for \catala an unusually simple and straightforward
interoperability scheme: direct source code generation to virtually any
programming language. This solution is generally impractical, requiring
considerable workarounds to reconcile semantic mismatches between target and
source, along with a significant run-time support library. In the case of
\catala, however, our intentional simplicity makes this “transpiling” scheme
possible.

Indeed, the final intermediate representation of the \catala compiler is a pure
and generic lambda calculus operating over simply-typed values. By re-using
standard functional compilation techniques such as closure conversion~\cite{minamide1996typed},
we claim that it is possible to compile \catala to any programming language
that has functions, arrays, structures, unions, and support for exceptions. We also believe that
a more complex version of the compilation scheme presented in \sref{lambdacalculus}
would remove the need for exceptions (in favor of option types), but leave this as future work.

The run-time required for generated programs only has to include an infinite precision
arithmetic library (or can default to fixed-sized arithmetic and floats) and a calendar
library to compute the days difference between two dates, taking leap years
into account. We demonstrate this with the OCaml backend of the \catala compiler,
which amounts to 350 lines of compiler code and 150 lines of run-time code (excluding
the \li+zarith+~\cite{zarith} and \li+calendar+~\cite{calendar} libraries).
Merely compiling to OCaml already unlocks multiple target environments,
such as the Web, via the
\li+js_of_ocaml+ compiler~\cite{vouillon2014bytecode}. We thus effortlessly bring
\catala to the Web.

\section{Putting \catala to Work}

The solid formal and technical foundations of \catala would be quite useless
if the language was not fit for its target audience: lawyers and legal
expert systems programmers. We claim that the design process of \catala as well
as our comprehensive code co-production process proposal maximizes the potential
for adoption by professionals. To support this claim, we report early user
study results and demonstrate an end-to-end use case with
the computation of an important French social benefit.

\subsection{Interacting with Lawyers}
\label{sec:lawyerinteraction}

\catala's design has been supervized and influenced by lawyers since its inception.
Indeed, the project started out of Sarah Lawsky's insight on the logical structure
of legal statutes~\cite{lawskykeynote, lawsky2017, lawsky2018, lawsky2020form}.
As well as providing the formal base building block of \catala, lawyers also
reviewed the syntax of \catala, choosing the keywords and providing insights
counter-intuitive to programmers, such as the \li+rule+/\li+definition+
distinction of \sref{rulesdefinitions}.

We also conducted a careful analysis of the production process of legal expert
systems. We found that in France, administrative agencies always use a V-shaped
development cycle for their legal expert systems. In practice, lawyers of the
legal department take
the input set of legal statutes and write a detailed natural language
specification, that is supposed to make explicit the different legal interpretations
required to turn the legal statute into an algorithm. Then, legal expert systems
programmers from the IT department take the natural specification and turn it into code, often never
referring back to the original statute text.

Exclusive interviews conducted by the authors with legal expert systems programmers
and lawyers inside a high-profile French administration reveal that this
theoretical division of labor is artificial. Indeed, the natural language
specification often proves insufficient or ambiguous to programmers, which
leads to programmers having to spend hours on the phone with the lawyers to clarify
and get the algorithm right. Furthermore, the validation of the implementation
depends on lawyer-written test cases, whose number and quality suffer from
budget restrictions.

This insight suggests that a more agile development process associating
lawyers and programmers from the beginning would be more efficient. We claim
the \catala is the right tool for the job, since it allows lawyers and programmers
to perform pair programming on a shared medium that locally combines the legal
text as well as the executable code.

We do not expect lawyers to write \catala code by themselves. A number of frameworks
such as Drools~\cite{drools} are built on this promise. For our part, we believe
that software engineering expertise is needed to produce maintainable, performant,
high-quality code. Hence, we envision for lawyers to act as observers and
reviewers of the code production process, safeguarding the correctness with
respect to the legal specification.

We don't expect adoption difficulties from the programmers' side, since \catala
is morally a pure functional language with a few oddities that makes it
well-suited
to legal specifications. To assess our claim of readability by lawyers, we
conducted a small user study with $N=18$ law graduate students enrolled in the
Masters program \enquote{Droit de la création et numérique}
(Intellectual Property and Digital Law) at Université Panthéon-Sorbonne.
The students are anonymous recruits, enrolled in a course taught by a lawyer
friend of the project. The study was conducted during a 2-hour-long video
conference, during which the students were able to submit feedback in real
time thanks to an online form. None of the students had any prior programming experience;
this question was asked orally at the beginning of the session.

The methodology of the study is the
following: the participants were given a 30 min. presentation of \catala's
context and goals, but were not exposed to any program or syntax. Then,
participants were briefed during 15 min. about Section 121 and its first
paragraph (\sref{121a}) and received a copy of the corresponding
\catala code (\sref{metadata} and \sref{rulesdefinitions}).  Finally, the participants
were told the protocol of the study:
\begin{itemize}
  \item 10 min. for reading the \catala code of
Section 121, then fill the questionnaire listed in \tref{userquestions};
  \item 15 min. of collective Q\&A  with the
  \catala authors (over group video conference)
  about the \catala code of Section 121, then fill a second time the same questionnaire listed in \tref{userquestions}.
\end{itemize}

The participants and experimenters carried out the study according to the aforementioned protocol.
After all students had filled the questionnaire
for the second time, a short debriefing was conducted.

\begin{table}
  \centering
  \smaller
\begin{tabular}{lp{.85\columnwidth}}\toprule
  \#&Exact text of the question\\\midrule
  (1)&Do you understand the project? Is there anything that is unclear about it?\\
  (2)&Can you read the code without getting a headache?\\
  (3)&Can you understand the code?\\
  (4)&Can you link the code to the meaning of the law it codifies?\\
  (5)&Can you certify that the code does exactly what the law says and nothing more?
   If not, are there any mistakes in the code?\\
\bottomrule
\end{tabular}
\caption{Questions of the user study\label{table:userquestions}}
\end{table}

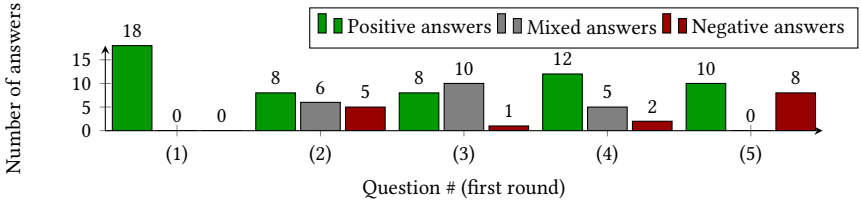
\begin{figure}
\centering
\smaller
\begin{tikzpicture}[font=\smaller]
  \begin{axis}[
      ybar,
      xlabel={Question \# (first round)},
      ylabel=Number of answers,
      width=0.8\columnwidth, %
      height=2.7cm,
      bar width=15pt,
      xmin=-0.5,
      xmax=4.5,
      ymin=0,
      axis lines=left,
      legend cell align=left,
      legend style={at={(1.05,1.45)},anchor=north east},
      nodes near coords={\pgfmathprintnumber[precision=0,fixed]{\pgfplotspointmeta}},
      legend columns=3,
      xtick={0,1,2,3,4},
      xticklabels={(1),(2),(3),(4),(5)},
  ]
  \addplot[fill=green!60!black]
      table[x=Question,y=Number,col sep=comma]
      {userstudyyesbefore.table};
  \addplot[fill=gray]
      table[x=Question,y=Number,col sep=comma]
      {userstudymixedbefore.table};
  \addplot[fill=red!60!black]
      table[x=Question,y=Number,col sep=comma]
      {userstudynobefore.table};
  \legend{Positive answers~,Mixed answers~,Negative answers}
  \end{axis}
  \end{tikzpicture}
\caption{Results of the first round of questions in the user study\label{fig:userstudyfirstround}}
\end{figure}

\begin{figure}
  \centering
  \smaller
  \begin{tikzpicture}[font=\smaller]
    \begin{axis}[
        ybar,
        xlabel={Question \# (second round)},
        ylabel=Number of answers,
        width=0.8\columnwidth, %
        height=2.7cm,
        bar width=15pt,
        xmin=-0.5,
        xmax=4.5,
        ymin=0,
        axis lines=left,
        legend cell align=left,
        legend style={at={(1.05,1.75)},anchor=north east},
        nodes near coords={\pgfmathprintnumber[precision=0,fixed]{\pgfplotspointmeta}},
        legend columns=3,
        xtick={0,1,2,3,4},
        xticklabels={(1),(2),(3),(4),(5)},
    ]
    \addplot[fill=green!60!black]
        table[x=Question,y=Number,col sep=comma]
        {userstudyyes.table};
    \addplot[fill=gray]
        table[x=Question,y=Number,col sep=comma]
        {userstudymixed.table};
    \addplot[fill=red!60!black]
        table[x=Question,y=Number,col sep=comma]
        {userstudyno.table};
    \legend{Positive answers~,Mixed answers~,Negative answers}
    \end{axis}
    \end{tikzpicture}
  \caption{Results of the second round of questions in the user study\label{fig:userstudysecondround}}
  \end{figure}
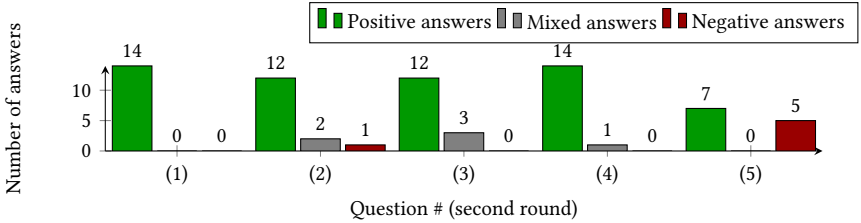

The full answers of the participants to all the questions of both rounds
are available in the artifact corresponding to this paper \cite{merigoux_denis_2021_4775161}.
The answers given by the participants in free text were interpreted as positive, negative
or mixed by the authors. \fref{userstudyfirstround} and \fref{userstudysecondround}
show the results for the first and second fillings of the questionnaire by the
participants.

These early results, while lacking the rigor of a scientific user study,
indicate a relatively good reception of the literate programming paradigm
by lawyers. The significant increase in positive answers between the first
and second round of questions indicates that while puzzled at first, lawyers
can quickly grasp the intent and gist of \catala once given a minimal amount
of Q\&A time (15 minutes in our study). One participant to the study was
enthusiastic about the project, and contacted the authors later to join the
project and tackle the modeling of French inheritance law in \catala.

After investigation, we believe that the large number of negative answers for question (5) in the second
round could be explained by a lack of familiarity with the US Internal Revenue Code from the French
lawyers. Indeed, the wording of the question (\enquote{certify}) implies that
the lawyer would be confident enough to defend their opinion in court. We believe,
from deeper interactions with lawyers closer to the project, that familiarity
with the formalized law combined with basic \catala training could bring the
lawyers' confidence to this level.

We deliberately introduced a bug in the code shown to the lawyers in
this user study. The bug involved a $\leqslant$ operator replaced by $\geqslant$.
Of the 7 who answered \enquote{Yes} to (5) in the second round, 2 were able
to spot it, which we interpret to be a very encouraging indication that lawyers
can make sense of the \catala code with just a two-hour crash course.

\subsection{A Look Back to Section 121}

We have used Section 121 of the US Internal Revenue Code as a support for introducing
\catala in \sref{tutorial}. But more interestingly, this piece of law is also
our complexity benchmark for legal statutes, as it was deemed (by a lawyer
collaborator) to be one of the
most difficult sections of the tax code.
This reputation comes from its dense wording featuring various layers
of exceptions to every parameter of the gross income deduction.

We have so far formalized it up to paragraph (b)(4), which is approximately
15\% of the whole section and around 350 lines of code (including the text of
the law), but contain its core and most used exceptions. We include the result
of this effort in the artifact \cite{merigoux_denis_2021_4775161}.
The current formalization was done in four 2-hour sessions of pair programming between
the authors and lawyers specialized in the US Internal Revenue Code. Progress is relatively
slow because we consider in the process every possible situation or input that
could happen, as in a real formalization process. However, this early estimate
indicates that formalizing the whole US Internal Revenue Code is a completely reachable target
for a small interdisciplinary team given a few years' time.

While finishing the formalization of Section 121 is left as future work, we are
confident that the rest of the section can be successfully expressed
in \catala: the maze of exceptions is localized to (a) and
(b), and the rest of the limitations are just a long tail of special cases; with
our general design that supports arbitrary trees of exceptions in default logic,
this should pose no problem.

\subsection{Case Study: French Family Benefits}
\label{sec:familybenefits}

\sref{lawyerinteraction} argues that \catala is received positively by lawyers.
This is only half of the journey: we need to make sure \catala is also
successfully adopted by the large private or public organization where legacy
systems are ripe for a principled rewrite.
To
support our claims concerning the toolchain and interoperability scheme in
a real-world setting, we formalized the entire French family benefits
computation in \catala and exposed the compiled program as an OCaml library
and JavaScript Web simulator. The full code of this example can be found in
the supplementary material of this article, although it is written in French.

A crucial part of the French welfare state, family benefits are distributed
to households on the basis of the number of their dependent children. Created
in the early 1930's, this benefit was designed to boost French natality by
offsetting the additional costs incurred by child custody to families. Family
benefits are a good example of a turbulent historical construction, as the
conditions to access the benefits have been regularly amended over the
quasi-century of existence of the program. For instance, while family benefits
were distributed to any family without an income cap, a 2015 reform
lowered the amount of the benefit for wealthy households~\cite{modulationafcontre, modulationafpour}.

The computation can be summarized with the following steps. First, determine
how many dependent children are relevant for the family benefits (depending on their
age and personal income). Second, compute the base amount, which depends on the
household income, the location (there are special rules for overseas territories)
and a coefficient updated each year by the government to track inflation. Third,
modulate this amount in the case of alternating custody or social services
custody. Fourth, apply special rules for when a child is exactly at the age
limit for eligibility,
or when the household income is right above a threshold.
All of these rules are specified by 27 articles of the French Social Security
Code, as well as various executive orders.

\begin{figure}
  \centering
  \includegraphics[width=0.85\columnwidth]{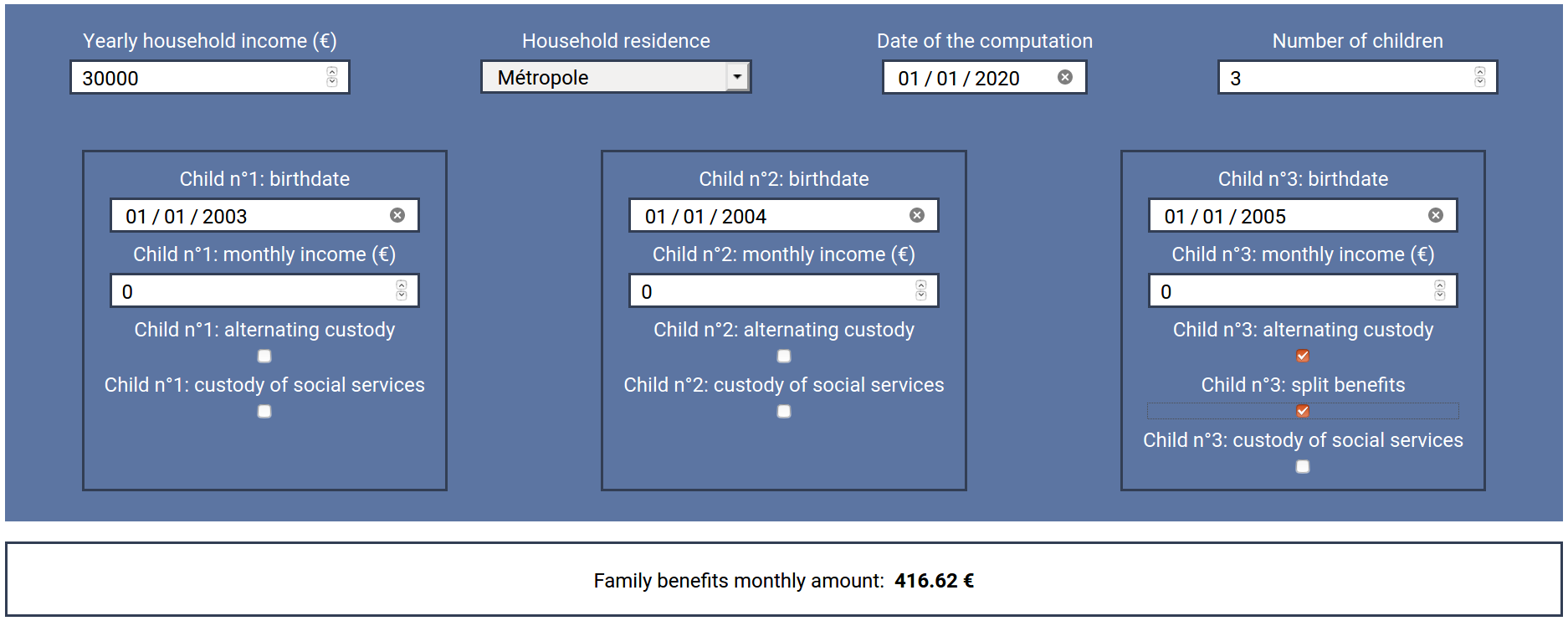}
  \caption{Screenshot of the Web family benefits simulator powered by \catala\label{fig:jssim}}
\end{figure}

The \catala formalization of this computation amounts to approximately 1,500
lines of code, including the text of the law. The code is split between 6 different
\li+scope+s featuring 63 \li+context+ variables and 83 \li+definition+s and
\li+rule+s. We believe these numbers to fairly representative of the authoring
effort required for a medium-size legal document.
Distributed as an OCaml library, our code for the computation of the French
family benefits is also used to power an online simulator (see \fref{jssim}).

After writing
the code as well as some test cases, we compared
the results of our program with the official state-sponsored simulator
\url{mes-droits-sociaux.gouv.fr}, and found no issue. However,
the case where a child is in the custody of social services was
absent from the official simulator, meaning we could not compare results for
this case. Fortunately, the source code of the simulator
is available as part of the OpenFisca software project~\cite{shulz2019logiciel}.
The \href{https://github.com/openfisca/openfisca-france/issues/1426}{OpenFisca
source file corresponding to the family benefits},
amounts to 365 lines of Python. After close inspection of the OpenFisca code,
a discrepancy was located with the \catala implementation. Indeed,
according to article L755-12 of the Social Security Code, the income cap
for the family benefits does not apply in overseas territories with single-child
families. This subtlety was not taken into account by OpenFisca, and was fixed
after its
\href{https://github.com/openfisca/openfisca-france/issues/1426}{disclosure by
the authors}.

Formalizing a piece of law is thus no different from formalizing a piece of
software with a proof assistant. In both cases, bugs are found in existing
software, and with the aid of a formal mechanization, can be reported and fixed.

\section{Conclusion \& Related Work}

\catala follows a long tradition of scholarly works that aim
to extract the logical essence of legal statutes, starting as early as
1924~\cite{dewey1924logical}. To provide some context, we compare our work
with two seminal articles in the field.

In his visionary 1956 article, Allen~\cite{allen1956symbolic} notes that symbolic logic can be used
to remove ambiguity in the law, and proposes its use for a wide range of
applications: legal drafting, interpretation, simplification and comparison.
Using custom notations that map transparently to first-order logic, Allen
does not provide an operational tool to translate law into formalism but rather
points out the challenges such as law ambiguity and rightfully sets the limits
of his approach, stating for instance that in generality, \enquote{filling of
gaps in legislation by courts cannot and should not be entirely eliminated}.
Interestingly, he also manually computes a truth table to prove that two
sections of the US Internal Revenue Code are equivalent.

The vision laid out by Allen is refined in 1986 by Sergot \emph{et al.}~\cite{nationalityact}.
This article narrows the range of its formalism to statutory law (as opposed to case law),
and focuses on the British Nationality Act, a statute used to determine whether
a person can qualify for the British nationality based on various criteria.
Co-authored by Robert Kowalski, this works features the use of Prolog~\cite{prolog}
as the target programming language, showing the usefulness of declarative
logic programming for the formalization task. However, the work acknowledges
a major limitation concerning the expression of negation in the legal text, and
points out that \enquote{the type of default reasoning that the act prescribes
for dealing with abandoned infants is nonmonotonic}, confirming the later
insights of Lawsky~\cite{lawsky2018}. A major difference with \catala
is the absence of literate programming; instead, Sergot \emph{et al.} derived
a synthetic and/or diagram as the specification for their Prolog program.

However, the line of work around logic programming never took hold in
the industry and the large organizations managing legal expert systems.
The reasons, various and diagnosed by Leigh~\cite{Leith2016}, mix the
inherent difficulty of translating law to code, with the social gap between
the legal and computer world. As a reaction, several and so far unsuccessful
attempts were made to automate the translation using natural language processing
techniques~\cite{pertierra2017towards, holzenberger2020dataset}. Others
claim that the solution is to lower the barriers to the programming world
using low-code/no-code tools, so that lawyers can effectively code their own
legal expert systems~\cite{morris2020spreadsheets}.

The main recent research direction around the formalization of law is spearheaded
by optimistic proponents of computational law~\cite{genesereth2015computational},
promising a future based on Web-based, automated legal reasoning by autonomous
agents negotiating smart contracts on a blockchain-powered network~\cite{
  hvitved2011contract, scoca2017smart, he2018spesc, zakrzewski2018towards}.

By contrast, we focus on the challenges related to maintaining existing
legal expert systems in large public or private organizations, providing essential
services to millions of citizens and customers. \catala aims to provide an
industrial-grade tool that enables close collaboration of legal and IT
professionals towards the construction of correct, comprehensive and performant
implementations of algorithmic statutory law.

The wide range of applications imagined by Layman in 1956 has yet to be
accomplished in practice. With its clear and simple semantics, we hope for
\catala formalizations of statutes to provide ideal starting point for future
formal analyses of the law, enabling legal drafting, interpretation,
simplification and comparison using the full arsenal of modern formal methods.

\begin{acks}
We want to thank first the lawyers at the heart of the \catala project:
Sarah Lawsky and Liane Huttner. Theirs insights and continued collaboration
were invaluable for the success of this endeavor. We also thank Pierre-Évariste
Dagand for its useful advice of encoding the default logic partial order into
a syntactic tree; this trick helped simplify a lot the formalization, without
loss of generality with respect to legislative texts.

This work is partially supported by the
\grantsponsor{ERC}{European Research Council}{https://erc.europa.eu/} under
the CIRCUS (\grantnum{ERC}{683032}) Consolidator Grant Agreement.
\end{acks}

\bibliographystyle{ACM-Reference-Format}
\bibliography{local,../../verifisc.bib}

\end{document}